\documentclass[a4paper,11pt,final]{article}
\pdfoutput=1 
\usepackage{jheppub} 
\usepackage[T1]{fontenc} 
\usepackage[utf8]{inputenc}
\usepackage{bm}
\usepackage{siunitx}
\usepackage{slashed}
\usepackage{showlabels}

\newcommand{\rew} {\mathrm{Re}\, \omega}
\newcommand{\imw} {\mathrm{Im}\, \omega}
\DeclareMathOperator{\Tr}{Tr}
\renewcommand{\Re}{\operatorname{Re}}
\renewcommand{\Im}{\operatorname{Im}}

\title{Freeze-in and freeze-out generation of lepton asymmetries after baryogenesis in the $\nu$MSM}

\author[a]{S. Eijima,}
\author[b]{M. Shaposhnikov,}
\author[b]{I. Timiryasov}

\affiliation[a]{ICRR, The University of Tokyo, Kashiwa, Chiba 277-8582, Japan}
\affiliation[b]{Institute of Physics, Laboratory for Particle Physics and Cosmology,\\
\'{E}cole Polytechnique F\'{e}d\'{e}rale de Lausanne, CH-1015 Lausanne, 
Switzerland}

\emailAdd{eijima@icrr.u-tokyo.ac.jp}
\emailAdd{Mikhail.Shaposhnikov@epfl.ch}
\emailAdd{Inar.Timiryasov@epfl.ch}

\abstract{The $\nu$MSM---an extension of the Standard Model by three relatively light singlet Majorana fermions $N_{1,2,3}$---allows for the generation of lepton asymmetry which is several orders of magnitude larger than the observed baryon asymmetry of the Universe. The lepton asymmetry is produced in interactions of $N_{2,3}$ (with masses in the GeV region) at temperatures below the sphaleron freeze out $ T \lesssim 130$ GeV and can enhance the cosmological production of dark matter (DM) sterile neutrinos $N_1$ (with the mass of the keV scale) happening at $T \sim 200$ MeV due to active-sterile neutrino mixing. 
This asymmetry can be generated in freeze-in, freeze-out, or later in decays of heavy neutral leptons.
In this work, we address the question of the magnitude of the late-time asymmetry  (LTA) generated by the heavy neutral leptons $N_{2,3}$  during their freeze-in and freeze-out, leaving the decays for later work. We study how much of this asymmetry can survive down to the lower temperatures relevant for the sterile neutrino DM creation. We find that this LTA could result in the production of a sizeable fraction of dark matter.  We also examine a role played by magnetic fields and the Abelian chiral anomaly in the generation of LTA, not accounted for in the previous studies. We argue that the production of LTA  can be increased significantly and make an estimate of the influence of this effect.
}

\begin{document} 
\maketitle
\flushbottom

\section{Introduction} 
\label{sec:introduction}
Despite its remarkable success, the Standard Model (SM) of particle physics fails to explain neutrino oscillations, the origin of matter-antimatter asymmetry of the Universe, and the nature of Dark Matter (DM). It has been suggested in refs.~\cite{Asaka:2005an,Asaka:2005pn} that all these shortcomings of the SM can be simultaneously addressed in its minimal extension with three singlet Majorana fermions,  the $\nu$MSM. 
The lightest of the three, $N_1$, is the DM particle with mass of the keV scale~\cite{Dodelson:1993je,Shi:1998km,Abazajian:2001nj,Asaka:2006nq,Laine:2008pg,Shaposhnikov:2020aen}. The other two, $N_2$ and $N_3$ are responsible  for both the active neutrino masses via the see-saw mechanism~\cite{Minkowski:1977sc,Yanagida:1979as,GellMann:1980vs,Mohapatra:1979ia,Schechter:1980gr,Schechter:1981cv} and the generation of baryon asymmetry of the Universe (BAU) through coherent oscillations of heavy neutrinos~\cite{Akhmedov:1998qx,Asaka:2005pn}.  
The latter ones are called heavy neutral leptons, HNLs in short. Notably, even after the freeze-out of sphalerons the $N_2$ and $N_3$ can keep producing lepton asymmetry~\cite{Shaposhnikov:2008pf,Canetti:2012kh}, which we refer to as late-time lepton asymmetry (LTA).

This asymmetry is crucial for the resonantly enhanced mechanism of sterile neutrino DM production~\cite{Shi:1998km}. If the concentration of DM is zero\footnote{\label{other_mechanisms}This is not necessarily the case. As has been found recently in ref.~\cite{Shaposhnikov:2020aen}, DM sterile neutrinos
can be created after inflation by universal four-fermion interaction in Einstein-Cartan gravity. Similar conclusions can be reached if the $\nu$MSM is supplemented by higher dimensional operators~\cite{Bezrukov:2011sz}.}
at $T\simeq100$~GeV, the resonant production seems to be the only option in the $\nu$MSM framework, since the astrophysical X-ray bounds on active-sterile neutrino mixing and mass bound from structure formation rule out (see, e.g.~\cite{Boyarsky:2018tvu}) the non-resonant production mechanism of \cite{Dodelson:1993je}. 
The LTA has to be quite large, of order $10^{-5}$ in units $n/s$, where $n$ is a number density and $s$ is the entropy density. This asymmetry is generated after the freeze-out of the sphalerons so that its value doesn't contradict the observed BAU. To be relevant, it has to survive until the temperatures around $\sim 200$~MeV, when the resonant production of DM takes place~\cite{Shi:1998km,Abazajian:2001nj,Asaka:2006nq,Laine:2008pg}.

Baryogenesis in the $\nu$MSM has attracted considerable attention from both theoretical  (an incomplete list of related refs.~\cite{Shaposhnikov:2006nn,Shaposhnikov:2008pf,Canetti:2010aw,Asaka:2010kk,Anisimov:2010gy,Asaka:2011wq,Besak:2012qm,Canetti:2012vf,Drewes:2012ma,Canetti:2012kh,Shuve:2014zua,Bodeker:2014hqa,Abada:2015rta,Hernandez:2015wna,Ghiglieri:2016xye,Hambye:2016sby,Hambye:2017elz,Drewes:2016lqo,Asaka:2016zib,Drewes:2016gmt,Hernandez:2016kel,Drewes:2016jae,Asaka:2017rdj,Eijima:2017anv,Ghiglieri:2017gjz,Eijima:2017cxr,Antusch:2017pkq,Ghiglieri:2017csp,Eijima:2018qke,Ghiglieri:2018wbs,Ghiglieri:2019kbw,Bodeker:2019rvr,Ghiglieri:2020ulj,Klaric:2020lov,Domcke:2020ety})  and experimental (see, e.g. \cite{Liventsev:2013zz,Aaij:2014aba,Artamonov:2014urb,Aad:2015xaa,Khachatryan:2015gha,Antusch:2017hhu,CortinaGil:2017mqf,Izmaylov:2017lkv,Mermod:2017ceo,Drewes:2018gkc,Ballett:2019bgd,Sirunyan:2018mtv,SHiP:2018xqw,Boiarska:2019jcw,Bolton:2019pcu,NA62:2020mcv,Tastet:2020tzh}) sides.    One of the reasons for such an interest is the testability of the model in the current and planned experimental facilities, such as  LHC~\cite{Sirunyan:2018mtv,Boiarska:2019jcw,Aad:2019kiz,Wulz:2019lsz},  NA62~\cite{NA62:2020mcv,Tastet:2020tzh,Drewes:2018gkc}, SHiP~\cite{Alekhin:2015byh,SHiP:2018xqw}, MATHUSLA~\cite{Curtin:2018mvb}, CODEX-b~\cite{Gligorov:2017nwh}, FASER~\cite{Feng:2017uoz,Kling:2018wct}, and ANUBIS~\cite{Hirsch:2020klk}.

On the contrary, the study of the LTA generation has been performed only in a few works, ref.~\cite{Canetti:2012kh,Canetti:2012vf,Ghiglieri:2019kbw,Ghiglieri:2020ulj}. In fact, {\em if one accepts that both  BAU and resonantly produced DM are the consequence of the $\nu$MSM see-saw Lagrangian~\cite{Minkowski:1977sc,Yanagida:1979as,GellMann:1980vs,Mohapatra:1979ia,Schechter:1980gr,Schechter:1981cv}},  the requirement of the generation of LTA big enough to explain the sterile neutrino DM abundance is the most restrictive  one~\cite{Canetti:2012kh,Canetti:2012vf,Ghiglieri:2020ulj}.  Given the potential of the forthcoming experiments---especially SHiP---to study a large portion of the parameter space below B-meson mass,  it is important to clarify if the HNLs responsible for both BAU and LTA production are within the experimental reach.

\begin{figure}[h]
    \centering
    \includegraphics[width=\textwidth]{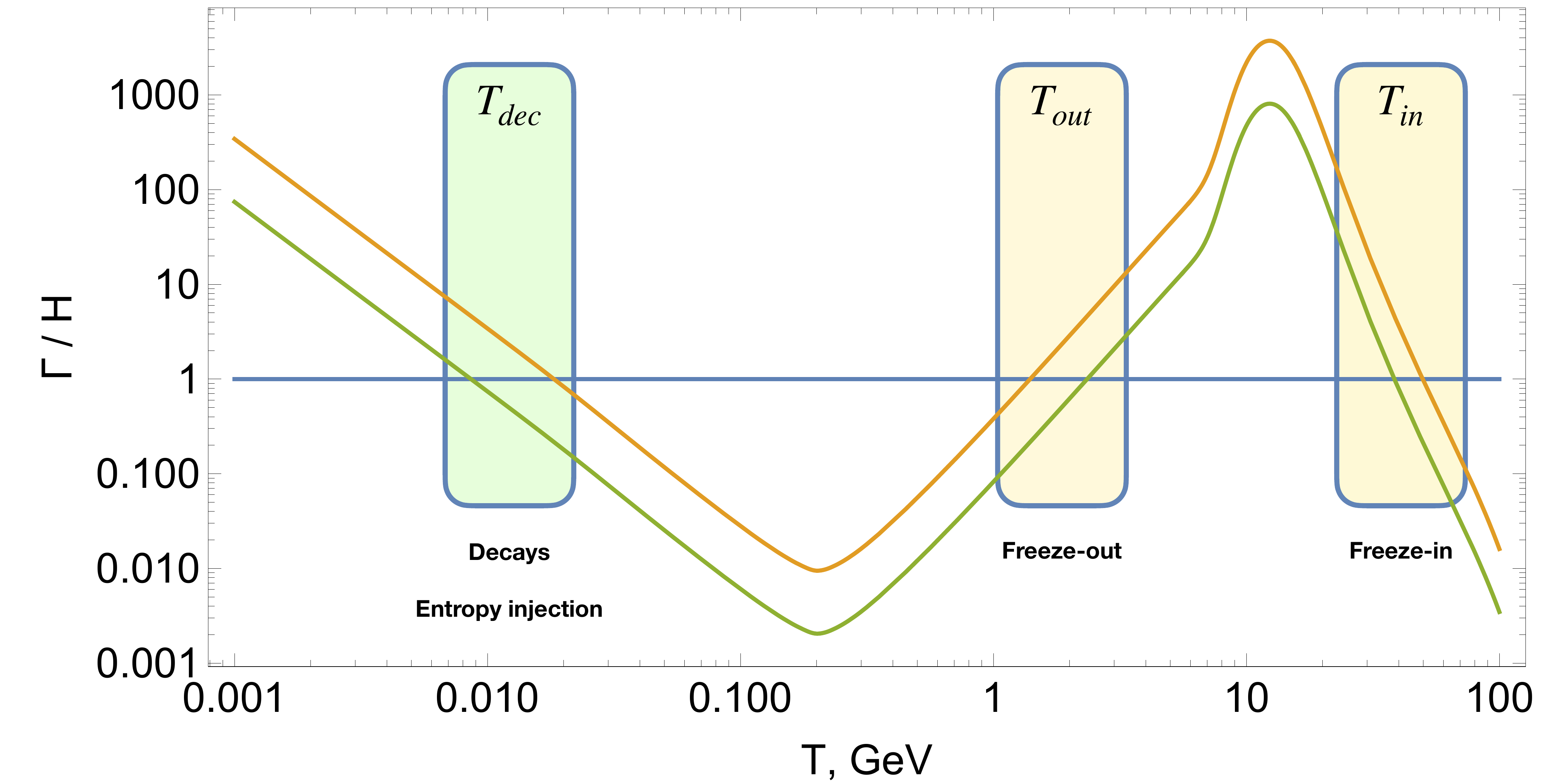}
    \caption{Interaction rates of two HNLs as functions of temperature (HNL mass is $1$~GeV, see figure~\ref{fig:rates} for more details). In this and the other similar plots time goes from right to left. There are three characteristic temperatures at which the HNL rates cross the Hubble rate and large asymmetry can in principle be generated: $T_{in}$, $T_{out}$, and $T_{dec}$. In this work we consider asymmetry generation around $T_{in}$ and $T_{out}$.}
    \label{fig:sketch}
\end{figure}
Let us briefly summarize a possible scenario of the evolution of the Universe within the $\nu$MSM. Right after inflation, the baryon and lepton numbers of the Universe, as well as the number of HNLs, may well be zero, and we will assume in the present paper that this is indeed the case~\cite{Bezrukov:2008ut} (see, however, ref.~\cite{Shaposhnikov:2020aen} and footnote~\ref{other_mechanisms}). The baryon asymmetry of the Universe is produced in a set of processes including coherent oscillations of HNLs,  exchange of the lepton number between the HNLs and active leptons, and anomalous sphaleron transitions \cite{Akhmedov:1998qx,Asaka:2005pn}.  The behaviour of HNLs after baryogenesis can be qualitatively understood from  figure~\ref{fig:sketch} showing their equilibration rates. It is seen that HNLs  enter in thermal equilibrium at $T_{in}$ which exceeds a few tens of GeV and subsequently go out of equilibrium at $T_{out} \sim 1$~GeV.

Owing to Sakharov non-equilibrium conditions, the lepton asymmetries can be generated at three instances: at freeze-in (temperature $T_{in}$),  at freeze-out  $T_{out}$, and during the HNL decays, at $T_{dec}$. The latter two possibilities have been considered in \cite{Shaposhnikov:2008pf,Canetti:2012kh}.  At the time these papers were written, the freeze-in LTA produced at $T_{in}$ was believed to be inessential. The arguments \cite{Shaposhnikov:2008pf,Canetti:2012kh,Ghiglieri:2016xye} were based on the fact that at $T_{out}<T<T_{in}$ the HNLs are in thermal equilibrium and thus all asymmetry which could have been produced at $T_{in}$ will be erased.

In fact, the situation happened to be more complicated \cite{Eijima:2017anv, Ghiglieri:2017gjz}. In spite of the fact that the HNL equilibration is much faster than the rate of the Universe expansion, some combination of lepton numbers and HNL asymmetries remains approximately conserved and thus is protected from wash-out. As a result, one can expect that part of LTA produced above $T_{in}$ survives till the low temperatures.~\cite{Eijima:2017anv, Ghiglieri:2017gjz}. So the question arises whether this asymmetry is enough for DM production. 
The present study aims at providing a quantitative answer to this question. 
Since the estimates of the interaction rates have changed considerably during the last few years, we also analyse the generation of the asymmetry during freeze-out. The possibility of production of large asymmetry during HNL decays has been recently demonstrated in ref.~\cite{Ghiglieri:2020ulj}. A systematic analysis of this possibility will be presented in a separate work~\cite{planned_paper}.

The paper is organised as follows. We start section~\ref{sec:generation_of_asymmetry} from a brief description of the main ingredients entering the calculation of asymmetries in the $\nu$MSM. In section~\ref{sub:hnl_decays_and_injection_of_entropy} we quantify the effect of entropy injection caused by the HNL decays. In section~\ref{sub:most_efficient_asymmetry_generation} we analyse the regime in which one can expect large LTA in the freeze-in; we also discuss the requirements for LTA generation during freeze-out. In section~\ref{sec:results} the previous considerations are put on the quantitative level by performing a scan of the parameter space. We use the kinetic equations averaged over the momentum. In section~\ref{sec:study_of_the_full_set_of_integro_differential_equations} we verify the findings of the previous section by solving the momentum-dependent equations. In section~\ref{sec:chiral_asymmetriy_and_magnetic_fields} we address whether the presence of Abelian chiral anomaly and possible asymmetry transfer into helical magnetic fields may play a role in LTA generation. We argue that such effects may indeed be  important and therefore should be systematically accounted for. In particular, the maximal value of the electron asymmetry reached during the evolution of the system could influence the final asymmetry. Therefore, in section~\ref{sec:maximal_electron_asymmetry} we perform a scan of the parameter space looking for the maximal electron asymmetry. 
In section~\ref{sec:possible_uncertainties_of_kinetic_equations} we discuss possible uncertainties of the state-of-the-art approach based on the kinetic equations which we adopt in this work. Finally, section~\ref{sec:discussion} contains our conclusions and outlook.

\section{Generation of asymmetry} 
\label{sec:generation_of_asymmetry}

In this section, we summarize the main ingredients required for the calculation of the BAU and LTA in the $\nu$MSM. All notations coincide with those of~\cite{Eijima:2018qke} to which we refer for further details.

The Lagrangian of the system is the well known see-saw 
one~\cite{Minkowski:1977sc,Yanagida:1979as,GellMann:1980vs,Mohapatra:1979ia,Schechter:1980gr,Schechter:1981cv}. 
We present it below in order to fix the notations.
In
the basis where charged lepton Yukawa couplings and the Majorana mass
term for the right-handed neutrinos are diagonal the Lagrangian can be written in the following form:
\begin{equation}
    \mathcal{L} = \mathcal{L}_{SM} + i \bar{\nu}_{R_I} \gamma^\mu \partial_\mu 
    \nu_{R_I}
    - F_{\alpha I} \bar{L}_\alpha \tilde{\Phi} \nu_{R_I} - \frac{M_{I J}}{2} 
    \bar{\nu}_{R_I}^c \nu_{R_J} + h.c.,
    \label{Lagr}
\end{equation}
where $\mathcal{L}_{SM}$ is the SM Lagrangian, $\nu_{R_I}$ are right-handed neutrinos, $I, J = 1, 2, 3$, $F_{\alpha I}$ is the matrix of 
Yukawa couplings, $L_\alpha$ are the left-handed lepton doublets, $\alpha = e, \mu, \tau$ and $\tilde{\Phi} = i\sigma_2 \Phi^*$,
 $\Phi$ is the Higgs doublet. 
Upon diagonalising the mass matrix following from~\eqref{Lagr} one finds three light mass eigenstates $\nu_i$ and three heavy mass eigenstates $N_I$. At the leading order of the see-saw mechansim $N_I = \nu_{R_I}$, whereas $\nu_{L \alpha}=U_{\alpha i} \nu_{i}+\Theta_{\alpha I} N_{I}^{c}$, $U_{\alpha i}$ is the PMNS matrix. 
In the last equation we have introduced the mixing angle $\Theta_{\alpha I} = v_0 F_{\alpha I}/M_I$, where $v_0=174$~GeV is the Higgs vacuum expectation value, see, e.g. ref.~\cite{Asaka:2011pb}.

The lightest right-handed neutrino is the dark matter candidate. The values of its Yukawa coupling constants are significantly constrained by astrophysical observables. As a consequence, the contributions to the 
active neutrino masses are negligibly small~\cite{Asaka:2005an}. Therefore, only two heavier HNLs provide neutrino masses and participate in the generation of both BAU and LTA. In what follows, we will limit our consideration to these two heavier HNLs, $N_2$ and $N_3$.
The only input from the lightest sterile neutrino is the value of the LTA required to boost the DM production. This value is not unique and depends upon the active-sterile mixings and the mass of $N_1$. A thorough study of DM production can be found in ref.~\cite{Ghiglieri:2015jua}.

Not all choices of Yukawas are compatible with the measured masses and mixings of active neutrinos. A convenient way of accounting for the oscillation data is given by the Casas-Ibarra parametrisation~\cite{Casas:2001sr}. In this parametrisation all Yukawas consistent with the observed oscillation data are determined by the following 6 parameters: the common mass $M$ of two HNLs;  mass splitting $\Delta M$; two CP-violating phases of the PMNS matrix $\delta$ and $\eta$\footnote{In the case of two HNLs, the PMNS matrix contains only one independent Majorana phase. Using PDG conventions one can identify the phase $\eta$  with  $(\alpha_{21} - \alpha_{31})/2$ in the case of normal hierarchy and $\alpha_{21}/2$ in the case of inverted hierarchy.}; real and imaginary parts of a complex angle $\omega$. The real part of $\omega$ enters all expressions as $\exp(-i \,\rew)$. The imaginary part $\imw$ controls the size of the Yukawas, namely
$|F_{\alpha\, I}| \propto \exp(\imw)$ for large positive $\imw$ and $|F_{\alpha\, I}| \propto \exp(-\imw)$ for large negative $\imw$.

Description of the evolution of the system comprising two HNL species and the SM degrees of freedom is rather complicated. The reason is that the asymmetry production is a genuinely non-equilibrium phenomenon, and it involves many processes, such as scatterings and decays of HNLs, their coherent oscillations, transfer of the asymmetry to leptons and their back reaction,
and redistribution of the asymmetry among the SM degrees of freedom.
These processes can be systematically accounted for in the integro-differential kinetic equations~\cite{Asaka:2005pn,Shaposhnikov:2008pf} which have to be solved numerically.
In this work, we use the equations of ref.~\cite{Eijima:2018qke}.\footnote{These equations have been derived in the relativistic approximation. As a result, the Boltzmann suppression is not present in the terms proportional to $f_{\nu} (1-f_{\nu})$. We have confirmed numerically that this suppression does not change the results for the whole mass range considered here. A more general form of kinetic equations can be found in ref.~\cite{Klaric:2020lov}.} 
These equations provided a unified description of low-scale leptogenesis models, including both resonant leptogenesis and baryogenesis via oscillations~\cite{Klaric:2020lov}.
They can be written in terms of the matrix of densities $\rho_N$ of two HNLs ($\rho_{\bar{N}}$ for HNLs of opposite helicity) and the densities of the combinations
$\Delta_\alpha = L_\alpha - B/3$ for $\alpha = e, \mu, \tau$ which are not affected by the sphaleron processes.
\begin{subequations}
\begin{align}
i \frac{d n_{\Delta_\alpha}}{dt}
&= -  2 i \frac{\mu_\alpha}{T} \int \frac{d^{3}k}{(2 \pi)^{3}} \Gamma_{\nu_\alpha} f_{\nu} (1-f_{\nu})  \, 
    + i \int \frac{d^{3}k}{(2 \pi)^{3}} \left( \, \text{\text{Tr}}[\tilde{\Gamma}_{\nu_\alpha} \, \rho_{\bar{N}}]
    -  \, \text{\text{Tr}}[\tilde{\Gamma}_{\nu_\alpha}^\ast \, \rho_{N}] \right),\label{kin_eq_a0}
\\
i \, \frac{d\rho_{N}}{dt} 
&= [H_N, \rho_N]
    - \frac{i}{2} \, \{ \Gamma_{N} , \rho_{N} - \rho_N^{eq} \} 
    - \frac{i}{2} \, \sum_\alpha \tilde{\Gamma}_{N}^\alpha \, \left[ 2 \frac{\mu_\alpha}{T} f_{\nu} (1-f_{\nu}) \right],\label{kin_eq_b0}
\\
i \, \frac{d\rho_{\bar{N}}}{dt} 
&= [H_N^\ast, \rho_{\bar{N}}]
    - \frac{i}{2} \, \{ \Gamma_{N}^\ast , \rho_{\bar{N}} - \rho_N^{eq} \} 
    + \frac{i}{2} \, \sum_\alpha (\tilde{\Gamma}_{N}^\alpha)^\ast \, \left[ 2 \frac{\mu_\alpha}{T} f_{\nu} (1-f_{\nu}) \right],
\label{kin_eq_c0}
\end{align}
\label{kin_eq}\end{subequations}
where $f_\nu = \left( e^{k/T}+1 \right)^{-1} $ is the Fermi-Dirac distribution function of a massless neutrino in equilibrium, 
$\rho_N^{eq} = diag(1,1) \left( e^{E_N/T}+1 \right)^{-1}$  is the matrix of densities of HNLs in equilibrium and $E_{N}=\sqrt{k^{2} + M^{2}}\;$.
Chemical potentials to $\Delta_\alpha$ are related to the number densities as $\mu_\alpha = \omega_{\alpha \beta}(T) n_{\Delta_\beta}$,
where $\omega_{\alpha \beta}(T)$ is the (inverse) susceptibility matrix, see, e.g.~\cite{Ghiglieri:2016xye,Eijima:2017cxr}.
Notice that  equations \eqref{kin_eq} agree with linearised equations from refs.~\cite{Ghiglieri:2017gjz,Ghiglieri:2019kbw,Bodeker:2019rvr}.
All the processes listed above are encoded in the rates entering the kinetic equations. Computation of these rates poses a theoretical challenge on its own~\cite{Anisimov:2010gy,Ghiglieri:2017gjz}. Here we use the results of the state-of-the-art computations of ref.~\cite{Ghiglieri:2018wbs}.

Solving integro-differential kinetic equations is very time-consuming, so to allow for a scan of the parameter space we simplify them. 
Namely, we assume that the matrix of densities of the HNLs is proportional to the equilibrium one. This ansatz allows reducing the infinite system of integro-differential equations to a set of 11 ordinary differential equations with averaged rates, see ref.~\cite{Eijima:2018qke} for the details. 
However, one needs to keep in mind that this ansatz is rather ad hoc and can be justified only by solving the full system. We perform this task in section~\ref{sec:study_of_the_full_set_of_integro_differential_equations}, see also refs.~\cite{Asaka:2011wq, Ghiglieri:2017csp}. For the interesting parameter sets (leading to large LTA), we find that the asymmetry obtained from the averaged equations agrees with the accurate one within a factor of $\sim 1.3$.

At the end of this section, we introduce for convenience the so-called yields $Y_X$ 
\begin{equation}
  Y_X\equiv \frac{n_X}{s}, \quad n_X = \int\frac{dk^3}{(2 \pi)^3} \rho_X,
  \label{yield}
\end{equation}
where we use the entropy density $s$ as computed in refs.~\cite{Laine:2006cp,Laine:2015kra}.
These quantities are not affected by the expansion of the Universe without extra entropy injection.

\section{HNL decays and injection of entropy} 
\label{sub:hnl_decays_and_injection_of_entropy}

\begin{figure}[htb!]
    \centering
    \includegraphics[width=0.9\textwidth]{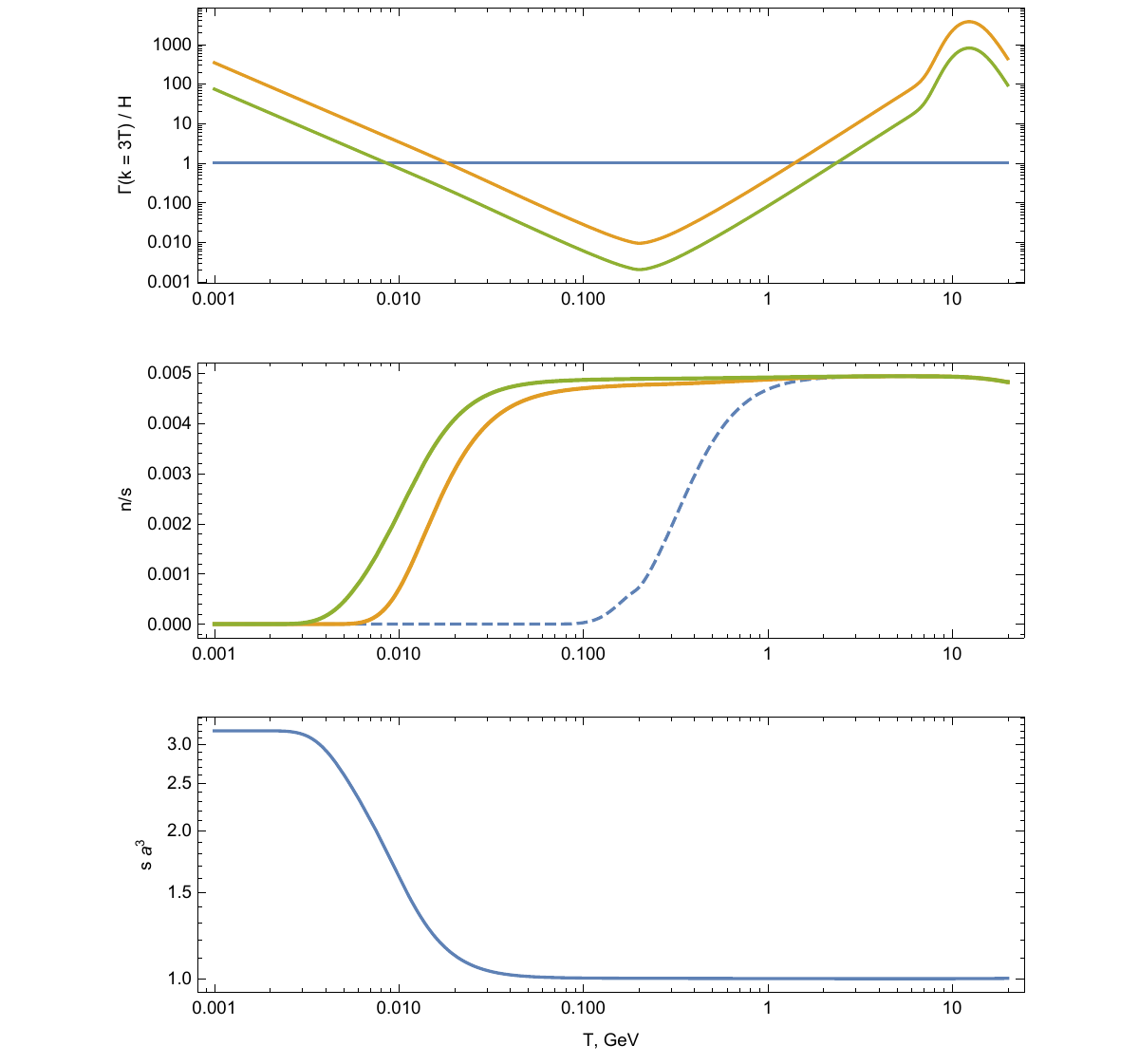}
    \caption{\emph{Upper panel:} the ratio $\Gamma_I/H$ as a function of temperature. Green and orange curves correspond to  $\left( \Gamma_N \right)_{22}$ and $\left( \Gamma_N \right)_{33}$. Horizontal line is placed at $\Gamma_I/H = 1$.
    \emph{Central panel:}  Actual number density of HNLs normalized to the entropy density  (solid lines) and the equilibrium one (dashed line). HNLs deviate from equilibrium when the rates cross the Hubble rate.
    \emph{Lower panel:} dilution factor $s\,a^3$ as a function of temperature. Out-of-equilibrium decays of HNLs increase the value of $s\, a^3$.
    The parameters are fixed to the values $M=1$~GeV, $\Delta M = 10^{-11}$~GeV, $\imw = 2\times10^{-3}$, $\rew = 0.5 \pi$, $\delta = 1.5 \pi$, and $\eta = 0.5 \pi$.} 
    \label{fig:rates}
\end{figure}

The HNLs that have participated in the generation of BAU and LTA eventually decay into the SM particles.\footnote{In this work, we do not consider CP violation and lepton asymmetry which can be generated in these decays.}  Slow decays of the HNLs eject additional entropy, thus diluting the otherwise conserved quantities, such as $Y_B$ or $n_{DM}/s$ \cite{Scherrer:1984fd,Asaka:2006ek}.
This effect can be accounted for directly in eqs.~\eqref{kin_eq}~\cite{Ghiglieri:2019kbw}. 
However, since the freeze-in asymmetry generation takes place well before the decays of HNLs, one can facilitate the numerics by computing the effect of the entropy injection separately.
The HNL decays take place at low temperatures and the corresponding rates can be approximated \cite{Canetti:2012kh}  by zero-temperature vacuum decay widths of the HNLs~\cite{Gorbunov:2007ak,Bondarenko:2018ptm}.\footnote{
Our results  slightly differ from those in ref.~\cite{Ghiglieri:2019kbw}. The reason is that we use the zero-temperature width of HNLs which accounts for all decay channels~\cite{Gorbunov:2007ak,Bondarenko:2018ptm}, whereas an effective number of flavours is used in~\cite{Ghiglieri:2019kbw}.}
The rates describing the HNLs decays read
\begin{equation}
    \left( \Gamma_N^{\rm \,dec} \right)_{II} = \frac{M_I}{E_N} \sum_{X_\alpha} \Gamma(N_I \to X_\alpha),
    \label{low_T_rates}
\end{equation}
where the sum goes over all allowed final states $X_\alpha$  and $\Gamma(N_I \to X_\alpha)$ is a partial decay width in the rest frame of an HNL~\cite{Bondarenko:2018ptm}.
Owing to the factor $M_I/E_N$, the decay rates~\eqref{low_T_rates} decrease at high temperatures.
 As an example, the total rates for $M = 1$~GeV HNLs 
are shown in the upper panel of figure~\ref{fig:rates}. 

One can expect from figure \ref{fig:rates} that the HNLs with $|\imw| \ll 1$ will deviate from  equilibrium at $T = \mathcal{O}(1)$~GeV. They will start to decay at temperatures around $\mathcal{O}(10)$~MeV when $\Gamma_I$, which at low temperatures are determined by the vacuum decays and do not explicitly depend on temperature, surpass the Hubble rate again. These decays cause extra entropy injection~\cite{Scherrer:1984fd,Asaka:2006ek} as shown in the lower panel of figure~\ref{fig:rates}.
In order to quantify this statement we follow the procedure of ref.~\cite{Ghiglieri:2019kbw} using the rates as in eq.~\eqref{low_T_rates}. We solve the corresponding kinetic equations from temperature $T_{in} = 5$~GeV till $T_{fin} = 0.1$~MeV numerically for different masses of the HNLs and compute the dilution factor $s a^3$, where $a$ is the scale factor.
As introduced in section~\ref{sec:generation_of_asymmetry}, $Y_X$ is a conserved quantity if there are no processes changing the number density of particle species $X$ and the entropy in the co-moving volume.
In fact, the baryon to entropy ratio $Y_B$ freezes out at $T<T_{sph}\simeq130$~GeV, and the generation of lepton asymmetries $Y_{\Delta_\alpha}$ also stops at temperatures below $\sim10$~GeV.
It is the entropy injection that diminishes both $Y_B$ and $Y_{\Delta_\alpha}$ subsequently. 
Since this dilution happens after HNLs cease  generating $Y_B$ and $Y_{\Delta_\alpha}$,
it is sufficient to compute $s\,a^3$ independently and then use the fact that  
$Y_X(T_{fin}) = Y_X(T_{in}) (s\, a^3)_{in} / (s\, a^3)_{fin}$.

\begin{figure}[h]
  \centering
  \includegraphics[width=0.6\textwidth]{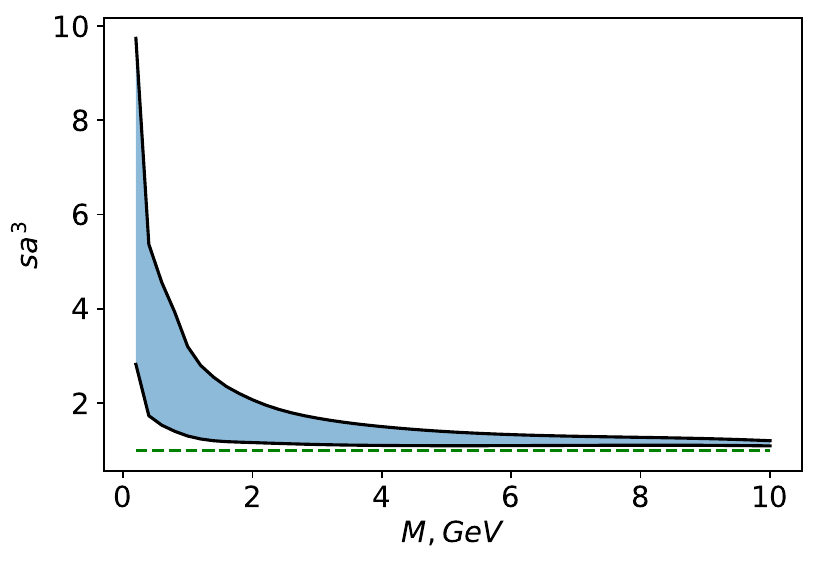}
  \caption{Dilution factor $s\,a^3$. All yields $Y_X =n_X/s$ are effectively divided by this factor.  The upper line corresponds to $\Im \omega = 0$: the smallest Yukawas, the long-lived HNLs, the maximal dilution.
   The lower line corresponds to $\Im \omega = 2$, in this case Yukawas are larger and the HNLs decay faster.}
  \label{fig:dilution}
\end{figure}

 The dilution factor for various values of $\Im \omega$ is shown in figure~\ref{fig:dilution}, in which we normalized $(s\, a^3)_{in} = 1$.
 It is large if the HNL decays are slow, that corresponds to small mixings and small masses. For a given mass the mixing takes its smallest value for $\imw = 0$.
Note that the upper black curve in figure~\ref{fig:dilution} should be treated with caution. Indeed, a very light HNL may be long-lived, thus
causing problems with Big-Bang nucleosynthesis~\cite{Dolgov:2000jw,Ruchayskiy:2012si,Gelmini:2020ekg,Sabti:2020yrt,Boyarsky:2020dzc}.


\section{The most efficient asymmetry generation} 
\label{sub:most_efficient_asymmetry_generation}

The dynamics described by system \eqref{kin_eq} depends on the parameters in a complicated way.
For example, as we have already mentioned, $ \exp \left( \imw \right) $ controls the size of the Yukawas and hence the size
of the damping rates $\Gamma_N$, $\Gamma_{\nu_\alpha}$, and the efficiency of the asymmetry transfer from HNL to neutrino sector and back. Therefore, large $|\imw|$ causes efficient asymmetry generation and, at the same time, its efficient wash-out. In this section, we qualitatively describe a regime in which large lepton asymmetry can be produced after the electroweak phase transition. The findings of this section are quantitatively confirmed in the next one.
 
\subsection{Asymmetry production during freeze-in} 
\label{sub:asymmetry_production_during_freeze_in}

In this subsection, we describe the freeze-in regime, which occurs at higher temperatures. Most of our findings are also applicable to the freeze-out generation of LTA, which we discuss in further detail in the next subsection.

\begin{figure}[h]
    \centering
    \includegraphics[width=0.6\textwidth]{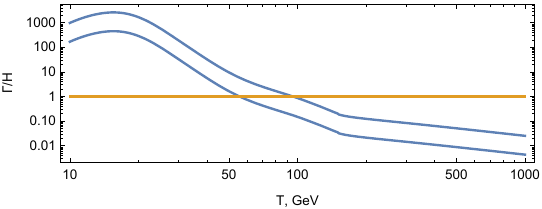}
    \includegraphics[width=0.6\textwidth]{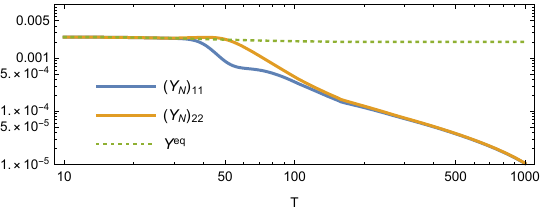}
    \includegraphics[width=0.6\textwidth]{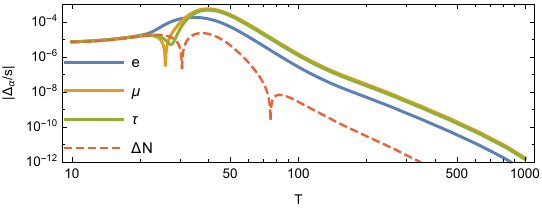}
    \includegraphics[width=0.6\textwidth]{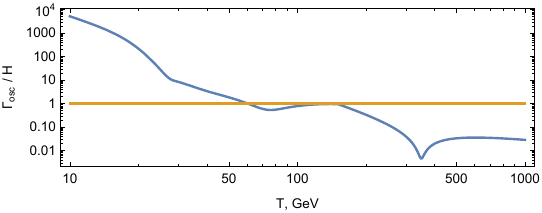}
    \caption{An example of generation of large LTA.
    \emph{First panel} displays two eigenvalues of the HNL equilibration rate $\Gamma_N$ divided by the Hubble rate $H$ as functions of temperature $T$.
    \emph{Second panel} shows the HNL yields as functions of $T$. The green dotted line indicates the equilibrium yield.  \emph{Third panel} shows the evolution of the lepton asymmetries (solid lines) and HNL asymmetry (dashed line)~\eqref{HNL_asym}. \emph{Fourth panel} shows the oscillation rate $\Gamma_{osc}$ divided by the Hubble rate $H$ (see eq.~\eqref{Gamma_osc}).
     The asymmetry is maximized if $\Gamma_{osc}/H\simeq1$ in the region where the eigenvalues of the rate $\Gamma_N$
    cross the Hubble. The model parameters are specified in eq.~\eqref{example_parameters}.}
    \label{fig:example}
\end{figure}

The major obstacle to large asymmetries at low temperatures is the wash-out. However, the wash-out is not always efficient. Since there are many different rates, let us illustrate the statement considering the following parameter set.
\begin{equation}
\begin{aligned}
    M & =2.0\text{~GeV}, \quad \Delta M = 0.983\times10^{-11}\text{~GeV},\quad \imw = -2.754\times 10^{-3}, \\
    \rew &= 0.551 \, \pi, \quad \delta = 0.993 \, \pi, \quad \eta = 1.479 \, \pi.
\end{aligned}
\label{example_parameters}
\end{equation}
This parameter set is quite representative since it results in large LTA.   
The HNL damping rates $\Gamma_N$ for the parameter set~\eqref{example_parameters}
 are shown in the upper panel of figure~\ref{fig:example}, in which the HNL rates cross the Hubble rate around $50-100$~GeV. 
 The other rates exhibit very similar patterns. 
Soon after $\Gamma_N/H$ becomes larger than $1$, HNL densities start to follow equilibrium line $Y^{eq}\simeq 0.0025$ (the second panel in figure~\ref{fig:example}). 
Let us, however, consider the total asymmetry in the HNL sector
\begin{equation}
\Delta_{N}=\left[\int \frac{d^{3} k}{(2 \pi)^{3}} \operatorname{Tr}\left(\rho_{N}-\rho_{\bar{N}}\right)\right].
\label{HNL_asym}
\end{equation}
The evolution of $\Delta_N/s$ is shown in the third panel of figure~\ref{fig:example} (red dashed curve). One can see that the asymmetry is much smaller than $Y^{eq}\simeq 0.0025$, but it is non-vanishing. In fact, the value $(\Delta_N/s)/Y^{eq}\sim 10^{-3}$ is surprisingly large given that the HNL rates exceed the Hubble rate by several orders of magnitude.

In order to understand why the asymmetry $\Delta_N$ is partially preserved during the period of equilibrium, it is instructive to 
rewrite the kinetic equations in the matrix form
\begin{equation}
    \frac{d n(T)}{dT} = A(T) n(T) + n^{S}(T),
    \label{matrix_form}
\end{equation}
where $n(T)$  is a column with $11$ real entries (3 lepton asymmetries $Y_{\Delta_\alpha}$, diagonal elements of hermitian matrices $Y_\pm$, and real and imaginary parts of the off-diagonal elements of 
$Y_\pm$),  $n^{S}(T)$ is a column of source terms associated with time derivative of $\rho_N^{eq}$. Above we introduced  $Y_+ = ( Y_N+Y_{\bar{N}} )/2 - Y^{eq}$ and $Y_- = Y_N-Y_{\bar{N}}$, see~\cite{Eijima:2018qke} for details. For the HNL masses we are considering here, the source term is irrelevant\footnote{Precisely speaking, the deviation from equilibrium caused by the expansion of the Universe could drive the asymmetry generation, see, e.g.~\cite{Klaric:2020lov}. However, the asymmetries generated in this way are at most of the order $\sim 10^{-7}$, i.e., several orders of magnitude smaller than the required ones.}
and the whole information is contained in the matrix $A(T)$.
The imaginary parts of the eigenvalues of $A(T)$ describe oscillations, whilst the real parts correspond to the production and wash-out of asymmetries. 

\begin{figure}[h]
    \centering
    \includegraphics[width=0.6\textwidth]{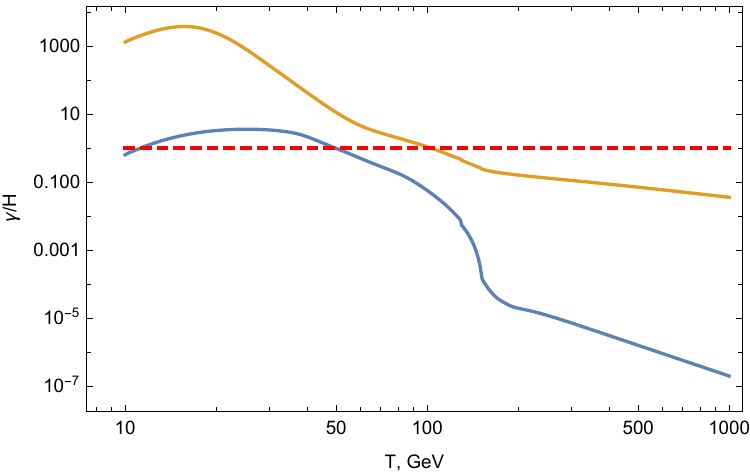}
    \caption{The real parts of the smallest (blue) and the largest (orange) eigenvalues of the matrix $A$ normalized to the Hubble rate as functions of temperature.}
    \label{fig:eigen}
\end{figure}

The real parts of the smallest and the largest eigenvalues of  $A$ as functions of temperature are shown in figure~\ref{fig:eigen}.
As one can see, the smallest eigenvalue exceeds the Hubble rate only in a relatively short temperature interval. 
This means that the linear combination of the variables corresponding to this eigenvalue is almost conserved. 
The weight coefficients of these linear combinations vary with temperature. 
We have performed a numerical scan with $M\leq 5$~GeV and found 
two combinations
\begin{equation}
    L_{\pm} \simeq \Delta_{N} \mp \sum_{\alpha} \Delta_{\alpha},
    \label{conserved_combination}
\end{equation}  
which are nearly conserved either at $T \gtrsim 150$~GeV ($L_-$) or $T\lesssim20$~GeV ($L_+$). In the intermediate region the linear combination corresponding to the smallest eigenvalue is more complicated and contains all $11$ variables $n(T)$ entering equation~\eqref{matrix_form}. The conditions for conservation of $L_-$ have been described in~\cite{Shaposhnikov:2008pf}. Both 
combinations~\eqref{conserved_combination} correspond to the approximately conserved quantum numbers identified in ref.~\cite{Eijima:2017anv} (notation in~\eqref{conserved_combination} is chosen to match that of ref.~\cite{Eijima:2017anv}). Indeed, as it has been shown in refs.~\cite{Eijima:2017anv} and~\cite{Ghiglieri:2017gjz}, the time derivative of $L_-$ is proportional to fermion number violating (helicity-conserving) rates which are suppressed in the symmetric phase. In turn, the time derivative of $L_+$ is proportional to the fermion number conserving rate~\cite{Eijima:2017anv} which becomes small at low temperatures. One particular consequence of the approximate conservation of $L_+$ 
is that $\Delta_N \simeq - \Delta_\alpha$ at low temperatures. The numerical solution of the kinetic equations confirms this observation, as can be seen in the third panel 
of figure~\ref{fig:example}.

We see that the presence of an almost conserved combination corresponding to the smallest eigenvalue of $A$ protects the asymmetry from being completely washed out. 
As we have already mentioned, the value of $|\imw|$ controls the size of Yukawa couplings. 
The values of Yukawas determine, in turn,
the eigenvalues of the matrix $A$. Therefore, if we are interested in the presence
of a small eigenvalue, $|\imw|$ should be close to zero.

Now once we have clarified how the asymmetry can survive, we need to understand how it can be generated. Thermally produced asymmetry is enhanced by the oscillations of the HNLs~\cite{Akhmedov:1998qx,Asaka:2005an}. This mechanism is the most efficient around the first few oscillations. The oscillation phase $\phi_{osc} = \int d t\, (E_3(T)-E_2(T))$ averaged over momentum can be approximated as
\begin{equation}
    \phi_{osc} \simeq \frac{\Gamma_{osc}}{H}, \quad \Gamma_{osc} = \frac{M \delta M(T)}{3\,T},
    \label{Gamma_osc}
\end{equation}
where we introduced the rate of oscillations $\Gamma_{osc}$. $H$ is the Hubble rate and 
$\delta M(T)$ is temperature-dependent physical mass difference.
It is also convenient to introduce
\begin{equation}
    T_{osc}(T) \simeq \left( M \delta M (T) M_0 /3 \right)^{1/3}.
    \label{Tosc}
\end{equation}
The asymmetry production is enhanced when $\Gamma_{osc}$ is close to the Hubble rate.
In terms of~\eqref{Tosc} this happens when $T\simeq T_{osc}(T)$.


The physical mass difference of the HNLs $\delta M$ is given by
\begin{equation}
    \delta M = \frac{E_N}{2 M} \Delta \lambda,
    \label{dM_def}
\end{equation}
with $\Delta \lambda = \lambda_2 - \lambda_1$, where $\lambda_{1,2}$ are the eigenvalues of the effective Hamiltonian $H_N$ entering eqs.~\eqref{kin_eq}. 
In order to clarify the parametric dependence of the physical mass difference, we present the corresponding expression at
zero temperature (in the case of normal hierarchy of neutrino masses, NH for short).
\begin{equation}
\left( \delta M \right)^2  \simeq \Delta M^{2}+\Delta M\left(m_{3}-m_{2}\right) \cos (2 \Re \omega)+\frac{1}{4}\left(m_{3}-m_{2}\right)^{2}+\mathcal{O} \left( \left( \frac{\Delta M}{M} \right)^2  \right) ,
\label{dM_zeroT}
\end{equation}
where $m_{2,3}$ are the masses of active neutrinos and 
${m_3-m_2 \simeq \sqrt{m_{atm}^2}\simeq 5 \times 10^{-11}~\text{GeV} }$\footnote{In the framework of the $\nu$MSM the mass of the lightest active neutrino is negligibly small~\cite{Asaka:2005an}.}. 
The similar expression in the case of inverted hierarchy (IH) can be obtained by replacing ${m_{2} \rightarrow m_{1}, m_{3} \rightarrow m_{2}}$.

The terms containing active neutrino masses at zero temperature in \eqref{dM_zeroT} originate from the Yukawa interactions in the Lagrangian. If there are no cancellations between Majorana $\Delta M$ and Yukawa contributions to the physical mass difference, the physical mass splitting is larger than the atmospheric mass difference.
In this case, the oscillation temperature is bounded from below by
\begin{equation}
    T_{osc} \gtrsim 180~\text{GeV} \left( \frac{M}{1 \text{GeV}} \right)^{1/3}.
    \label{bound_Tosc}
\end{equation}
This temperature exceeds $T_{sph}\simeq 130$~GeV. Therefore, to generate large lepton asymmetry which would not contradict the measured value of BAU, one has to lower $T_{osc}$. 
This can be achieved if $\rew \simeq \pi/2$. In this case the second term in \eqref{dM_zeroT} is negative and can offset two other terms.
The fine-tuning issues related to the cancellation of different terms and radiative corrections in~\eqref{dM_zeroT} have been discussed in refs.~\cite{Roy:2010xq,Canetti:2012kh}.

Note that CP-violation in the oscillations of HNLs vanishes when $\rew$ is $\pi/2$ exactly. We discuss this in greater detail in appendix~\ref{sec:late_time_asymmetries_and_inverted_hierarchy}. It is hard to generate large lepton asymmetry with $\rew = \pi/2$ because only CP violation in backreactions to leptons contributes. Figure~\ref{fig:scatter_plot} demonstrates this dependence on $\rew$.

The last requirement is that the first oscillation has to take place when the rates are large, but HNLs are not in equilibrium yet, like in the lower panel of figure~\ref{fig:example}. This can be achieved for each value of $M$ by a specific choice of $\Delta M, \; \rew, \; \imw$ and two phases. In order to identify such choices we have performed a scan of the parameter space which is described in section~\ref{sec:results}.

\subsection{Asymmetry production during freeze-out} 
\label{sub:asymmetry_production_during_freeze_out}
Another deviation from equilibrium when the asymmetry can potentially be generated happens during freeze-out, at $T\sim T_{out}$, see figure~\ref{fig:sketch}. 
Inspecting the rates, one can see that the wash-out of the asymmetry is no longer 
an issue. Indeed, by the very definition, both HNL and neutrino densities and asymmetries remain constant after the freeze-out.
In fact, the previous study~\cite{Canetti:2012kh} has indicated that the rates should be larger compared to the freeze-in case to allow for asymmetry generation. As we discussed in the previous subsection, the thermally produced asymmetry needs to be enhanced by the HNL oscillations. The enhancement is the most effective during the first few oscillations, i.e. ${T_{osc}\sim T_{out}}$. 
One can see from eq.~\eqref{Tosc} that low $T_{osc}$ requires very small physical mass difference $\delta M$. This is possible only provided a specific choice of the parameters. In particular, $\rew$ should be very close to $\pi/2$ and 
${\Delta M \simeq  (m_3-m_2)/2}$. Since $T_{out}\ll T_{in}$, the cancellations between different terms in~\eqref{dM_zeroT} should be much more delicate.

Summarising the points above, we can identify the requirements for the freeze-out generation of the LTA.
\begin{itemize}
    \item In contrast to the freeze-in case, the parameter $\imw$ needs not to be close to zero since the asymmetry remains constant after the freeze-out.
    Note that $T_{out}$---the temperature at which the HNL rates cross the Hubble rate---is determined by the rates and thus by the value of $\imw$.
    \item The oscillation rate $\Gamma_{osc}$ should be close to the Hubble rate around $T_{out}$, which is possible if $\delta M (T)$ is very small. This implies the requirements $\rew\simeq \pi/2$ and $\Delta M \simeq \sqrt{m_{atm}^2}/2$ (see eq.~\eqref{dM_zeroT}).
\end{itemize}
As in the freeze-in case, massive numerical scans are required to identify the optimal parameter sets. We discuss our numerical approach in the next section.




\section{Late-time lepton asymmetries} 
\label{sec:results}
In this section, we present the main results obtained through the numerical solution of equations \eqref{kin_eq} integrated over momentum. First, we briefly discuss the resonant mechanism of sterile neutrino dark matter production and the required late-time lepton asymmetries. Next, we describe the numerical procedure and present the main results and comment on them.

\subsection{Required LTA} 
\label{sub:required_lta}
As we have already mentioned, thermal DM production~\cite{Dodelson:1993je} via mixing with active neutrinos cannot be responsible for $100\%$ of DM abundance~\cite{Asaka:2006nq} given the astrophysical X-ray bounds on the active-sterile mixing. The large lepton asymmetry, if present at temperatures ${T_{\rm prod}\sim 200}$~MeV, can boost the DM production~\cite{Shi:1998km}. 
Computation of the asymmetry that is needed for producing the correct abundance of DM have been performed in refs.~\cite{Shi:1998km,Abazajian:2001nj,Asaka:2006rw,Asaka:2006nq,Laine:2008pg,Ghiglieri:2015jua}. The recent analysis of ref.~\cite{Bodeker:2020hbo} confirmed the findings of ref.~\cite{Ghiglieri:2015jua}.
Moreover, the results depend upon the flavour structures of the neutrino Yukawa couplings and different types of pre-existing lepton asymmetries. 
Here we follow ref.~\cite{Ghiglieri:2015jua}, which uses $M_{DM}=7.1$~keV as a benchmark. 
According to table 1 of ref.~\cite{Ghiglieri:2015jua}, the minimal initial asymmetry at $T= 4$~GeV yielding the correct dark matter abundance is $Y_{\nu_e}=Y_{\nu_\mu}=Y_{\nu_\tau} \simeq 11\times 10^{-6}$.
Note that all lepton asymmetries are equal at low temperatures owing to the presence of the two types of the rates~\cite{Eijima:2017anv}.
This translates into 
\begin{equation}
    Y_L \equiv \sum_\alpha Y_{\Delta_\alpha} \simeq 66\times 10^{-6},
\label{required_total}
\end{equation}
where we have defined the total lepton asymmetry $Y_L$.\footnote{The total asymmetry as it defined here also contains the BAU contribution. However, since $Y_B\ll Y_L$ in all interesting cases, this contribution can be safely neglected.}
Theoretical errors, mainly coming from hadronic uncertainties are expected at the level $10-20 \%$~\cite{Ghiglieri:2015jua}.

Note that the value \eqref{required_total} is not taking into account the effect of the entropy dilution discussed in section~\ref{sub:hnl_decays_and_injection_of_entropy}. If the two HNLs are sufficiently light and feebly coupled, their slow out-of-equilibrium decays which take place at $T\ll T_{\rm prod}$ inject additional entropy and, as a consequence, the otherwise conserved quantities, like $Y_B$ or $Y_{DM}$ become smaller. We will address this effect below.


\subsection{Numerical procedure and the freeze-in results} 
\label{sub:numerical_procedure_and_the_main_results}

Now our aim is to clarify whether the asymmetry at level \eqref{required_total} can be indeed generated via the mechanism described in section~\ref{sec:generation_of_asymmetry}. To answer this question, we have performed a scan of the parameter space. Equations~\eqref{kin_eq} were solved numerically using the Fortran code based on LSODE~\cite{LSODE}, see ref.~\cite{Eijima:2018qke} for details. The parameters considered in the scan are listed in table~\ref{table_parameters}. The upper limit on the HNL mass is chosen to ensure that our kinetic equations account for all relevant processes. In order to consider heavier HNLs, more decay channels have to be added, see~\cite{Klaric:2021cpi} for details. Our results below show that 
the mass range considered in table~\ref{example_parameters} is sufficient to study the  dominant channel of the LTA generation.

First we have scanned over the broad range of the parameters and identified the most interesting region. In this region (dubbed ``large LTA'' in table \ref{table_parameters}) we performed more detailed scans.
\begin{table}[htb!]
\begin{center}
  \begin{tabular}{| c | c | c | c | c | c | c |}
    \hline
   $M$, GeV & $\log_{10} (\Delta M/\mbox{GeV})$  & $\imw$ & $\rew$ & $\delta$ &
    $\eta$ & range:\\ \hline
   $[0.1 - 30]$  & $[-17,-9]$  & $[-2,2]$ & $[0, 2\pi]$ &$[0, 2\pi]$ & $[0, 2\pi]$ &broad range\\ 
    \hline
   $[0.1 - 30]$  & $[-12,-10]$  & $[-0.2,0.2]$ & $[0.4\pi, 0.6\pi]$ &$[0, 2\pi]$ & $[0, 2\pi]$ &large LTA\\ 
    \hline
  \end{tabular}
\end{center}
\caption{\label{table_parameters} Parameters of the theory:
common mass; Majorana mass difference; imaginary and real parts of $\omega$;
Dirac and Majorana phases.
In the second line, we indicate the full ranges of these parameters that were considered in this work, whereas the third line corresponds to the more restricted region where the large LTA can be generated.}
\end{table}
It is interesting to note that even for $\delta = 0, \eta = 0$ large asymmetry can be generated.
Let us stress that sufficiently large LTA can only be generated in the case of the normal hierarchy of neutrino masses.

\begin{figure}[h]
    \centering
    \includegraphics[width=0.6\textwidth]{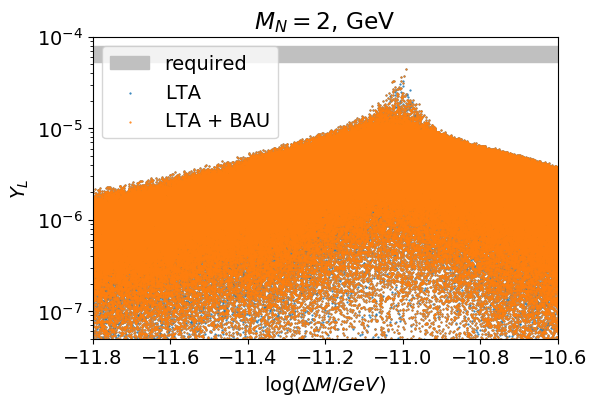}
    \includegraphics[width=0.6\textwidth]{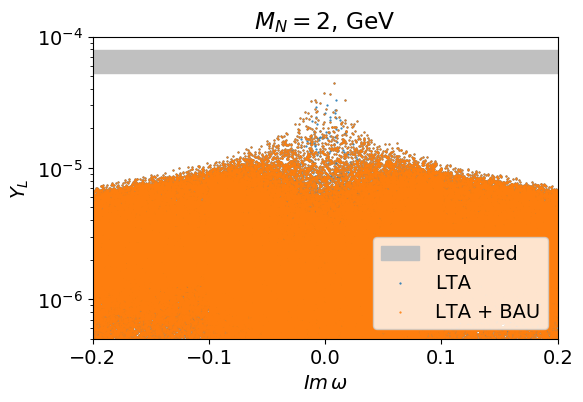}
    \includegraphics[width=0.6\textwidth]{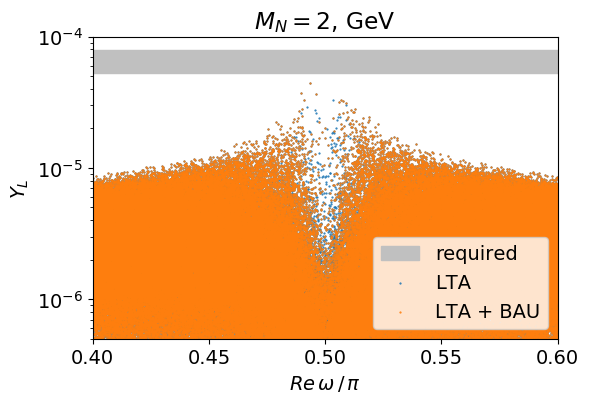}
    \caption{Dependence of the total lepton asymmetry on the parameters $\Delta M$ (upper panel), $\imw$ (middle panel), and $\rew$ (lower panel). Every point corresponds to a certain parameter set in the scan. Dependence on the Majorana and Dirac phases is rather flat, and therefore we do not show it here.}
    \label{fig:scatter_plot}
\end{figure}

\begin{figure}[h]
  \centering
  \includegraphics[width=0.6\textwidth]{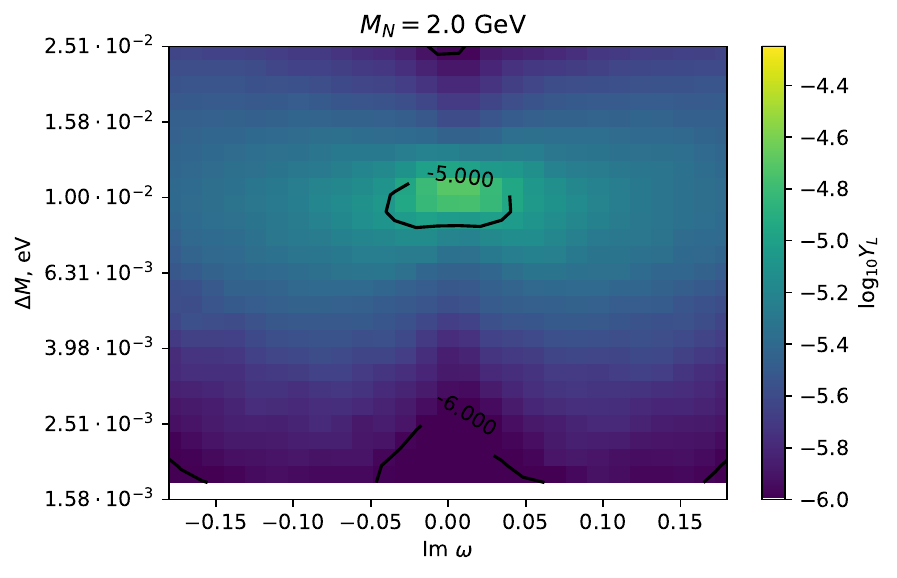}
  \includegraphics[width=0.6\textwidth]{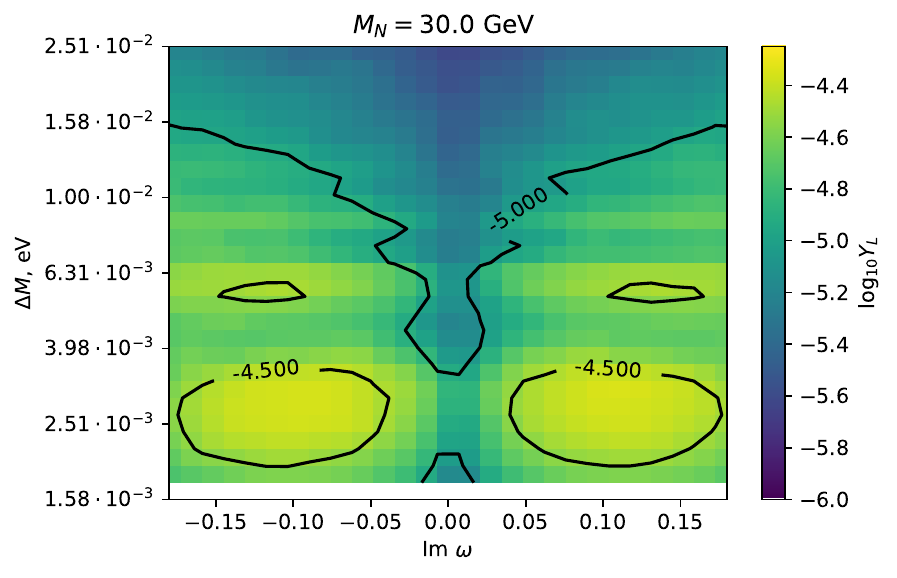}
  \caption{LTA as a function of $\imw$ and Majorana mass difference $\Delta M$.  Large value of LTA can be obtained for $\imw$ close to zero for
  sufficiently light HNLs ($M = 2$~GeV, upper panel). For heavier HNLs ($M = 30$~GeV, lower panel) another mechanism is operative and hence $\imw$ deviates from zero and mass splitting is smaller, see the main text for more detailed explanation.}
  \label{fig:dm_imw_orig}
\end{figure}

Our numerical analysis confirmed qualitative considerations of section~\ref{sub:most_efficient_asymmetry_generation} for $M<5$~GeV. Namely, $|\imw| \ll 1$ to ensure the smallness of the rates; $\rew \simeq \pi/2$ to allow for cancellations between Majorana and Higgs contributions to the physical mass difference, see figure~\ref{fig:scatter_plot}.
 LTA as a function of $\imw$ and Majorana mass difference $\Delta M$ is shown in figure~\ref{fig:dm_imw_orig}. 
As one can see from the figure, large asymmetry $ Y_L \equiv \sum_\alpha Y_{\Delta_\alpha}$ can be generated if $\imw$ is sufficiently small. 
In this region, the mixing angles between active neutrinos and HNLs denoted by $\Theta_{\alpha I}$ take their minimal values near the so-called "see-saw bound" $\imw=0$.
For example, the mixing summed over the lepton flavours and HNL generations reads
\begin{equation}
    |U|^2\equiv\sum_{\alpha, I} |\Theta_{\alpha I}|^2 = \frac{m_2+m_3}{M} \left[ \exp\left( 2\,\imw \right)  +\exp\left( - 2\,\imw \right) \right].
\end{equation}

\begin{figure}[h]
\centering
  \includegraphics[width=0.6\textwidth]{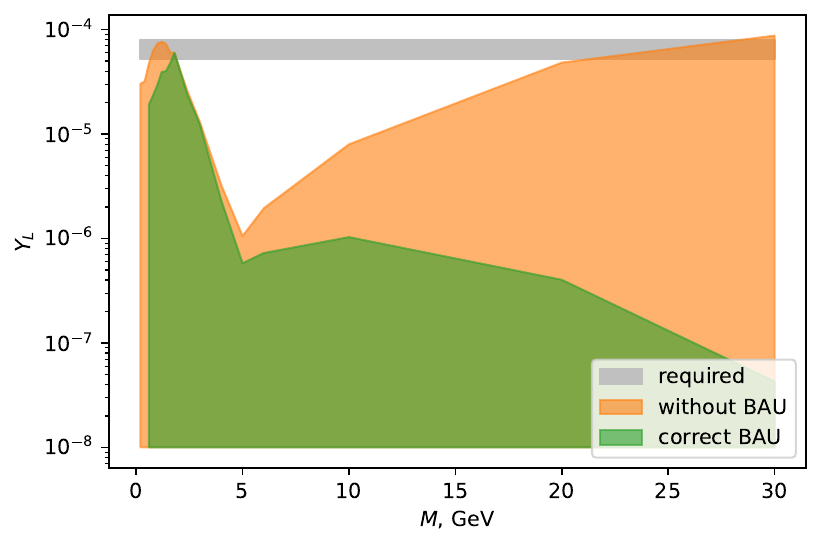}
  \includegraphics[width=0.6\textwidth]{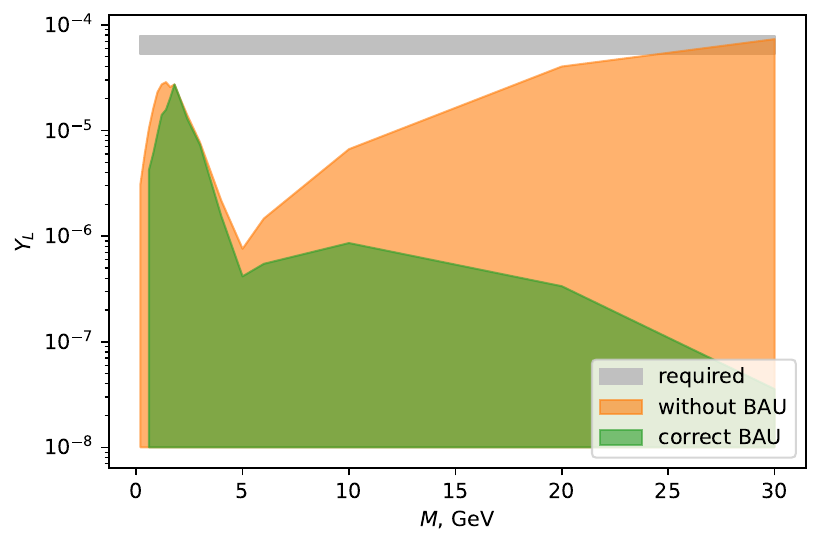}
  \caption{Total lepton asymmetry as a function of M. \emph{Upper panel:} at temperature $T=10$~GeV. \emph{Lower panel:} after dilution. The grey shaded region indicates the total lepton asymmetry which is needed for $N_1$ to compose $100\%$ of DM. Thickness of the curves indicates the possible uncertainties. Normal hierarchy. 
  }
  \label{fig:LTA}
\end{figure}

In figure~\ref{fig:LTA} we show the maximal value of the LTA as a function of HNL mass. The LTA can be indeed large for very light HNLs, but such HNLs cannot generate enough asymmetry by the time of the sphaleron freeze-out. Therefore the BAU is smaller than the observed value. Generated asymmetry can still be large  even if we require that the observed amount of BAU is generated, as is shown by the green shape in the figure.
\begin{figure}[h]
\centering
  \includegraphics[width=0.6\textwidth]{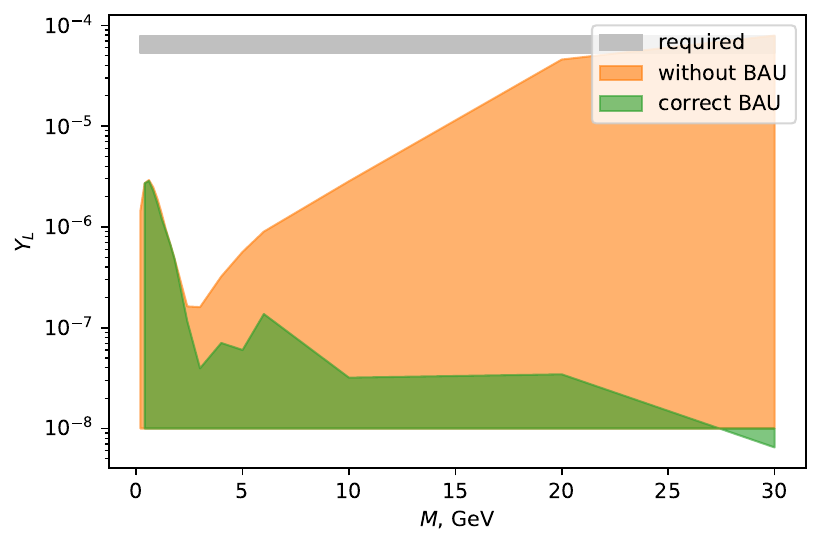}
  \includegraphics[width=0.6\textwidth]{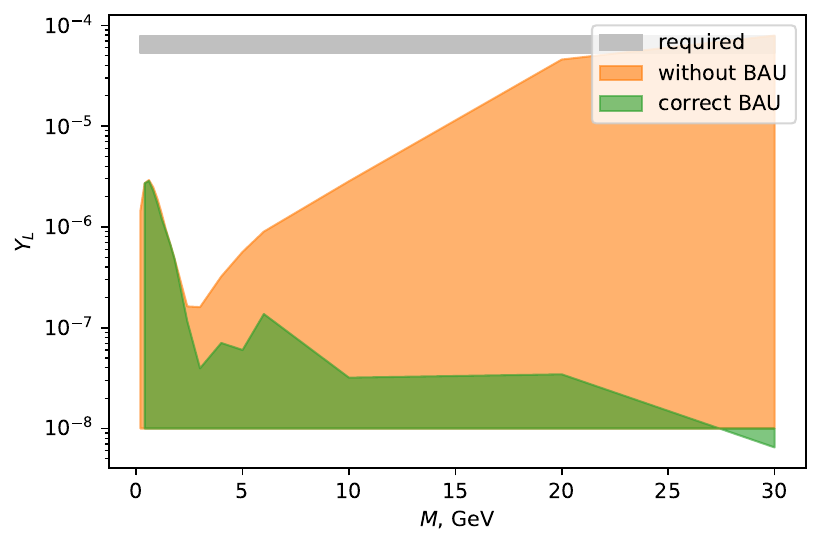}
  \caption{The same plots as figure~\ref{fig:LTA} but for the inverted hierarchy.}
  \label{fig:LTA_IH}
\end{figure}

The generated LTA boosts the DM production, which takes place at $T\sim 100$~MeV. 
Figure~\ref{fig:LTA} shows that the correct amount of DM can be produced for $M\simeq2$~GeV HNLs if all uncertainties are accounted for optimistically.
However, the decays of the very same HNLs that happen at $T\sim1$~MeV reduce both DM and BAU. To quantify the magnitude of dilution, we show the LTA divided by the dilution factor shown in figure~\ref{fig:dilution} as a function of mass. 
The result is shown in the lower panel of figure~\ref{fig:LTA}. Note that dilution takes place after the conversion of lepton asymmetry to DM, so the lower panel of figure~\ref{fig:LTA} shows an effective LTA.
Such effective description is valid since the DM abundance depends approximately linearly on the LTA in the region of interest~\cite{Ghiglieri:2015jua}.
The DM abundance computed with this effective LTA can be directly compared with the present-day value.
If all factor of $2$ uncertainties are  pushed in the direction which maximizes the generation of BAU and DM production, the $M\simeq2$~GeV HNLs can provide $\simeq 50 \%$ of the observed DM abundance. 

As it is clear from figure~\ref{fig:LTA_IH}, asymmetries are much smaller in the IH case. This can be understood by considering the detailed structure of the rates $\Gamma_N$ 
entering the kinetic equations~\eqref{kin_eq}. We address this in appendix \ref{sec:late_time_asymmetries_and_inverted_hierarchy}.

Below we qualitatively explain the behaviour observed in figures~\ref{fig:LTA} and~\ref{fig:LTA_IH} in the two distinct mass regions.

In the region $M<5$~GeV the qualitative picture of section~\ref{sub:most_efficient_asymmetry_generation} is confirmed. 
The rates increase with mass, and the wash-out becomes efficient above $M\simeq 2$~GeV.

\begin{figure}[h!]
    \centering
    \includegraphics[width=0.6\textwidth]{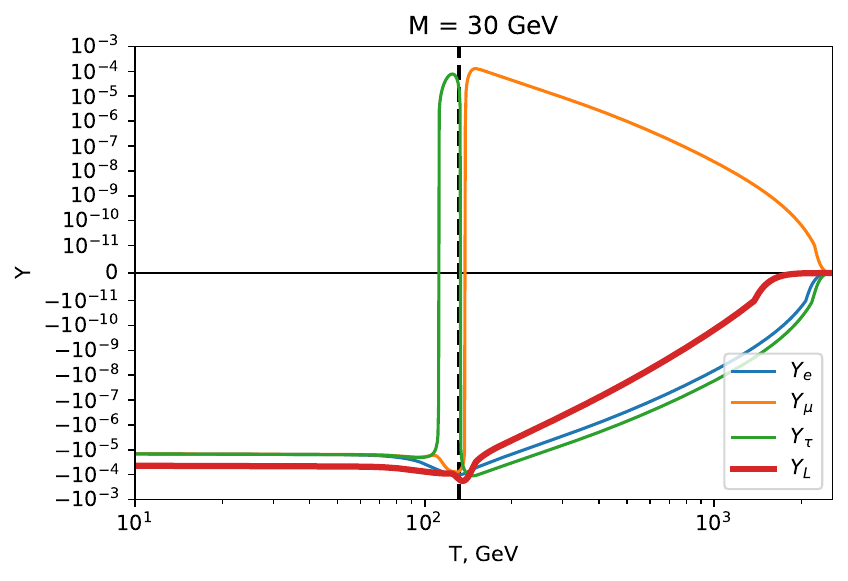}
    \caption{Total lepton asymmetry (red) and individual asymmetries (blue, orange, and green) as functions of temperature. The plot is logarithmic in both positive and negative directions of $Y$. The black dashed vertical line indicates the sphaleron freeze-out temperature $T_{sph}\simeq 130$~GeV where the baryon asymmetry freezes out. The total asymmetry and hence the baryon asymmetry are already large at $T_{sph}$. }
    \label{fig:evol}
\end{figure}

Another potentially interesting region revealed by the numerical scan is one of the relatively heavy HNLs with $M\gtrsim 10$~GeV.
Figures~\ref{fig:LTA} and \ref{fig:LTA_IH} exhibit that very large lepton asymmetry can be generated in this case. 
However, the production of such asymmetry due to the freeze-in of HNLs takes place around temperatures of sphaleron freeze-out $T_{sph}\simeq 130$~GeV. 
It means the BAU also turns out to be much larger than the observed value.
We demonstrate this in figure~\ref{fig:evol}.
Nevertheless, it is still interesting how the large asymmetry can survive until low temperatures. 
As one can see from figure~\ref{fig:dm_imw_orig}, relatively large values of $\imw$ and small mass splittings are preferred in the case of $M = 30$~GeV. This implies that the behaviour observed in figure~\ref{fig:evol} can be associated with an approximately conserved number which the pseudo-Dirac fermion formed by two HNLs involves~\cite{Blanchet:2009kk}.

\subsection{The freeze-out case} 
\label{sub:the_freeze_out_case}
\begin{figure}[h!]
    \centering
    \includegraphics[width=0.8\textwidth]{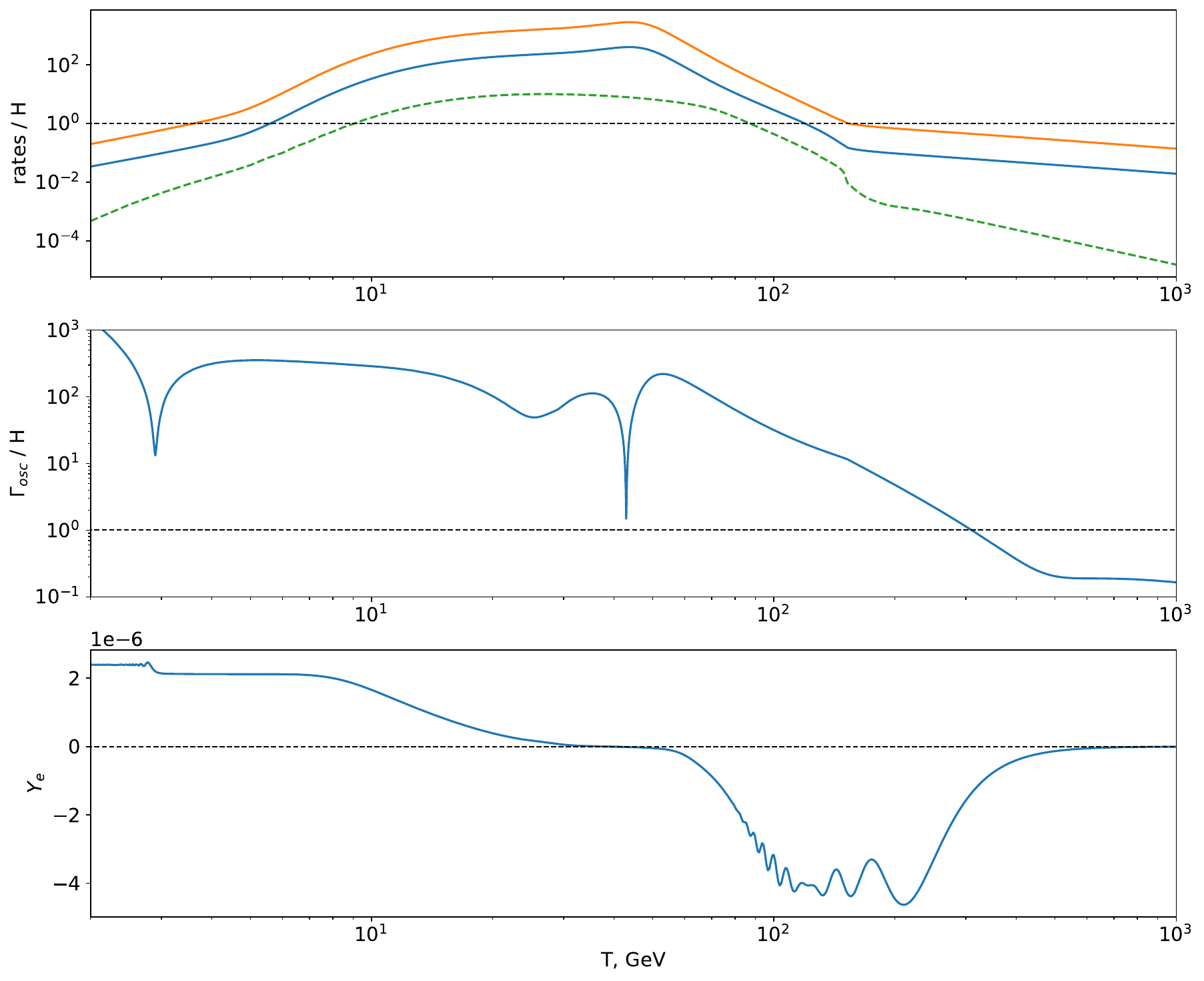}
    \caption{An example of the freeze-out production of the asymmetry. Upper panel: absolute values of the rates divided by the Hubble rate. Solid lines correspond to the eigenvalues of the rate $\Gamma_N$, while the green dashed line shows the smallest eigenvalue of the full matrix $A(T)$ (eq.~\eqref{matrix_form}). Middle panel: $\Gamma_{osc}/H$  as a function of temperature. Lower panel: the electron asymmetry (linear scale) as a function of time. One can see that the asymmetry is generated when the rates are smaller than the Hubble rate and when $\Gamma_{osc}/H$ is not much larger than 1.  The parameters are fixed to the values $M=10$~GeV, $\Delta M = 2.06\times10^{-11}$~GeV, $\imw = 0.2$, $\rew = 0.500001 \pi$, $\delta = 3.11 \pi$, and $\eta = 0.68 \pi$.}
    \label{fig:freezeout_example}
\end{figure}
Now we turn to the freeze-out case. An example of the freeze-out asymmetry production is shown in figure~\ref{fig:freezeout_example}. The upper panel of the figure shows two eigenvalues of the matrix $\Gamma_N(T)$ and the smallest eigenvalue of the full matrix $A(T)$ (see eq.~\eqref{matrix_form}). The middle panel shows the oscillation rate normalized to the Hubble rate
$\Gamma_{osc}/H$, eq.~\eqref{Gamma_osc}.
As one can see, this is a non-trivial function of temperature. The lower panel shows the evolution of the electron asymmetry. Let us examine this asymmetry in chronological order (time $t = M_0/(2 T^2)$ goes from right to left in the plot).
First, we see the freeze-in generation of asymmetry, which is the most efficient when  $\Gamma_{osc}/H \simeq 1$ and the rates are close to $H$ as well. Once the oscillation rates increases, the asymmetry production stops (the effect of the oscillations on $Y_e$ is visible). 
The asymmetry is washed out when the rates become large. 
Interestingly, when the oscillation rate briefly drops, the asymmetry changes its sign.
However, we did not try to find an analytic explanation of this feature.
Finally, the asymmetry builds up again when the rates decrease. Notice that the smallest eigenvalue (green dashed line in the figure) remains close to $1$. The asymmetry production is enhanced again at the point where the oscillation rate is small. 

The intricate dynamics described above requires numerical analysis.
Such an analysis is more challenging compared to the freeze-in one for several reasons. 
\begin{itemize}
    \item The asymmetry production takes place at lower temperatures, so the differential equations should be integrated over a longer time interval. Moreover, the HNL oscillations become very fast at low temperatures, rendering the equations even stiffer.
    \item The regions of the parameter space where the asymmetry generation is enhanced by the HNL oscillations are highly tuned. The reason for this tuning was explained in~\ref{sub:asymmetry_production_during_freeze_out}: both rates $\Gamma_{osc}$ and $\Gamma_N$ should be close to the Hubble rate.
\end{itemize}
As a result of these complications, the direct application of the numerical procedure described above is not sufficient. We can narrow down the regions of the scan using the following recipe.
First, for every HNL mass $M$ we identify the freeze-out temperature $T_{out}$ by solving $\Gamma_N(T_{out}) = H(T_{out})$. Since $\Gamma_N$ is a matrix, practically we equate its two eigenvalues to the Hubble rate and thus get two freeze-out temperatures, see figure~\ref{fig:freezeout_example}. We take $T_{out}$ to be lower of these two. Then we minimize the physical mass difference $\delta M(T)$ at $T  = T_{out}$. This specifies the values of the parameters $\Delta M$ and $\Re \omega$. These values give us an initial guess. The actual dynamics of the system is more complicated and cannot be described by just two temperatures, as can be seen already from figure~\ref{fig:freezeout_example}. Therefore, we perform a numerical scan in the vicinity of these values. Apart from this specific tuning of $\Delta M$ and $\Re \omega$, the scanning strategy is the same as in the freeze-in case.

\begin{figure}[h]
\centering
  \includegraphics[width=0.6\textwidth]{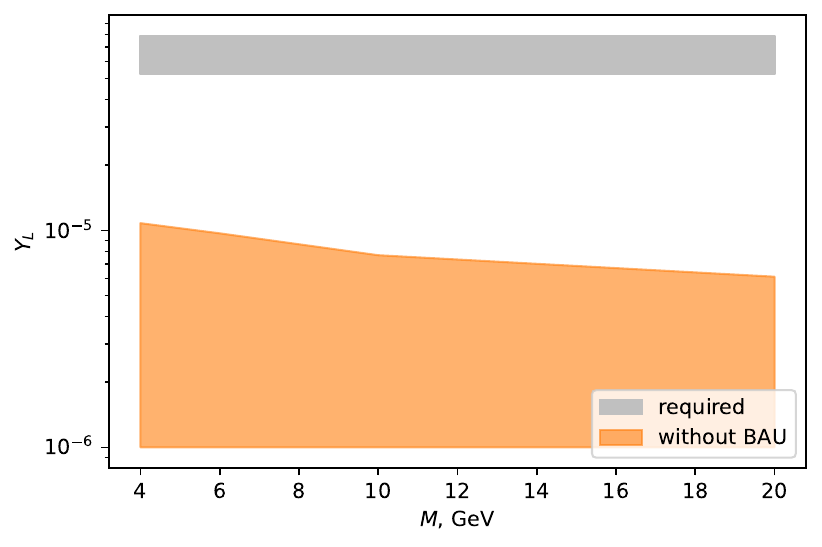}
  \caption{The total lepton asymmetry generated during freeze-out as a function of M.  Normal hierarchy. 
  }
  \label{fig:freezeout}
\end{figure}

\begin{figure}[h!]
    \centering
    \includegraphics[width=0.6\textwidth]{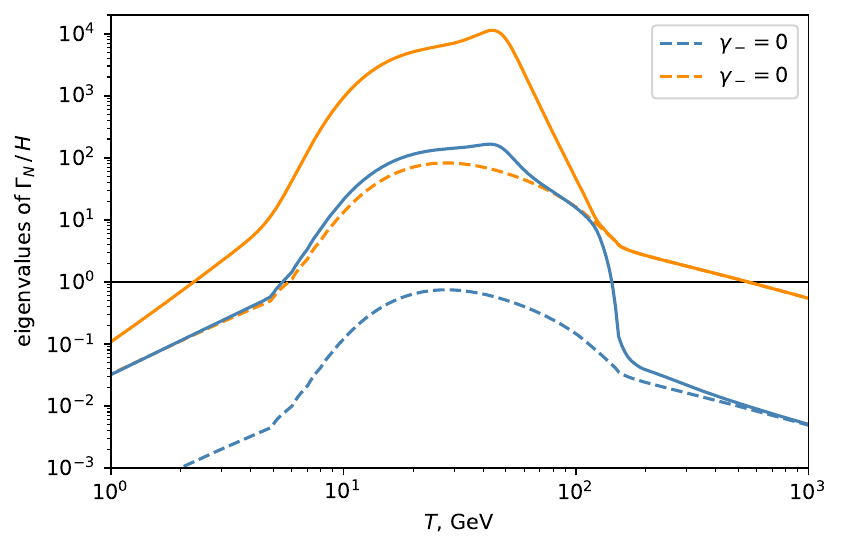}
    \caption{The eigenvalues of the rate $\Gamma_N$ divided by the Hubble rate. The solid line show the full rates used in this work. The dashed lines are obtained by artificially putting fermion number violating contributions $\gamma_-$ to zero. The $\gamma_-$ contributions were missing in some earlier studies. The parameters are fixed to the values $M=10$~GeV, $\Delta M = 2.06\times10^{-11}$~GeV, $\imw = 1.0$, $\rew = 0.500001 \pi$, $\delta = 3.11 \pi$, and $\eta = 0.68 \pi$.}
    \label{fig:fnc}
\end{figure}

The results of our scan are shown in figure~\ref{fig:freezeout}. 
As one can see, in our scan we have not found the parameter sets leading to sufficiently large LTA. This can seem surprising, especially given that the earlier studies, such as ref.~\cite{Canetti:2012kh} found that sufficient asymmetry can be generated in freeze-out. Note, however, that the presence of the fermion number violating contributions $\gamma_-$ to the rates (see~\cite{Eijima:2018qke} for details) has been realised only recently~\cite{Eijima:2017anv,Ghiglieri:2017gjz}. These contributions were missing in~\cite{Canetti:2012kh}.  The importance of $\gamma_-$ can be seen from figure~\ref{fig:fnc}.
As one can see, the change in the rates is drastic. Compared to the $\gamma_- = 0$ case, the freeze-out occurs later. Note also that in the $\gamma_- \neq 0$ case both eigenvalues of $\Gamma_N$ become larger than the Hubble rate, rendering the simultaneous tuning of $T_{osc}\sim T$ much more complicated. 

Note also that ref.~\cite{Canetti:2012kh} used a much weaker requirement for the asymmetry,
$Y_L\gtrsim8\times 10^{-6}$, which is almost one order of magnitude smaller than \eqref{required_total}. 
This difference is mainly due to the fact that~\cite{Canetti:2012kh} did not account for the 
distribution of asymmetry over different fermionic flavours. In addition, the update of~\cite{Laine:2008pg} made in~\cite{Ghiglieri:2015jua} resulted in increase of the necessary asymmetry.


Yet, we cannot completely exclude that there can exist very narrow regions of the parameter space leading to successful LTA generation, which have been missed by our scan. Finding such regions will require a dedicated analysis that goes beyond the scope of this work.


\subsection{Summary of the results of the numerical study} 
\label{sub:summary_of_the_results_of_the_numerical_study}

The main findings of our numerical study are the following.
\begin{itemize}
    \item Large ($\sim 10^{-5}$) LTA can be generated during freeze-in in the case of normal hierarchy (NH) of neutrino masses.
    \item The freeze-in generation means that all parameter sets leading to large LTA are concentrated close to the see-saw line $\imw \simeq 0$. In terms of the mixing between active neutrinos and HNLs summed over flavours and generations it means
    \begin{equation}
        |U|^2\simeq 6\times 10^{-11} \frac{\text{GeV}}{M}.
        \label{U2seesaw}
    \end{equation}
    \item If the theoretical uncertainty is accounted for in a generous way (that is, if the actual values of  BAU and LTA are twice larger than those obtained where), the total lepton asymmetry as large as $\simeq 6\times 10^{-5}$ and the observed value of BAU can be generated simultaneously.
    \item Slow decays of the HNLs \emph{after} generation of DM inject additional entropy thus diluting both DM abundance and BAU.
    \item We were able to find points where, after accounting for the entropy dilution, the $\nu$MSM can explain up to $\sim 50\%$ of $7~$keV DM. Note that the value of the required LTA depends on the mass and mixing angle of the DM. 
    \item The results above are obtained using the kinetic equations~\eqref{kin_eq}
    averaged over momentum with the rates as derived in refs.~\cite{Ghiglieri:2017gjz,Ghiglieri:2018wbs}. In the next section, we demonstrate that the averaging of the equations over momentum is a very good approximation. We consider other possible sources of uncertainty in section~\ref{sec:possible_uncertainties_of_kinetic_equations}. 
    Apart from these uncertainties, there might be new physical effects that are not accounted for in eqs.~\eqref{kin_eq}. In particular, we consider the processes related to the Abelian part of the anomaly in section~\ref{sec:chiral_asymmetriy_and_magnetic_fields}.
\end{itemize}
To sum up, if \emph{(i)} the dark matter abundance is vanishing at $T\sim 100$~GeV, \emph{(ii)} theoretical uncertainties do not exceed the values used in this section, and \emph{(iii)} effects not accounted for in our equations~\eqref{kin_eq} are not important, we can conclude that the lepton asymmetry generated in 
\emph{freeze-in} and \emph{freeze-out} is not enough to explain DM and BAU simultaneously.\footnote{Note that findings of ref.~\cite{Canetti:2012kh,Ghiglieri:2020ulj} show that both DM and BAU could be explained if the asymmetry is generated during  decays of HNLs. This mechanism allows for the production of $100 \%$ DM in both NH and IH cases if $\imw \sim \mathcal{O}(1)$ and the physical mass splitting is very tiny.} 
In the next section, we examine the validity of the assumption \emph{(iii)}.


\section{Study of the full set of integro-differential equations} 
\label{sec:study_of_the_full_set_of_integro_differential_equations}

The analysis so far was based on the averaged kinetic equations. In this section, we present the full set of the integro-differential kinetic equations in the form suitable for the numerical solution. We describe some details of the implementation and demonstrate that the results obtained using averaged and accurate equations agree very well.

Equations~\eqref{kin_eq} are introduced in flat space-time. In the expanding Universe the time derivative is replaced by ${
\frac{\partial}{\partial t} \rightarrow \frac{\partial}{\partial t}-H k^{i} \frac{\partial}{\partial k^{i}}}$. 
It is convenient to introduce 
\begin{subequations}
\begin{equation}
    z = \log(M/T), \quad y = k/T.
\end{equation}
The original equations are formulated in terms of $\rho_N$ and $\rho_{\bar{N}}$. We can rewrite them in terms of CP-even and CP-odd combinations $\rho_+ \equiv (\rho_N+\rho_{\bar{N}})/2 - \rho_N^{eq}$,  
${\rho_- \equiv \rho_N - \rho_{\bar{N}}}$. We also introduce ${\tilde{\omega}_{\alpha \beta} \equiv T^2 \omega_{\alpha \beta}}$. Now the equations~\eqref{kin_eq} can be rewritten as
\begin{align}
H\, \frac{\partial Y_{\Delta_\alpha}}{\partial z} = 
     &- 2 \, \tilde{\omega}_{\alpha \beta}(T) Y_{\Delta_\beta} \int \frac{d^3y}{(2\pi)^3} \Gamma_{\nu_\alpha} f_{\nu} (1-f_{\nu})  \nonumber\\
    &+ \frac{2 i\, T^3}{s} \int \frac{d^3y}{(2\pi)^3}\Tr \left( \rho_+ \, \Im  \tilde{\Gamma}_{\nu_\alpha}   \right)
    -\frac{T^3}{s} \int \frac{d^3y}{(2\pi)^3} \Tr \left( \rho_- \, \Re  \tilde{\Gamma}_{\nu_\alpha}  \right),\label{ke1}\\
H\, \frac{\partial \rho_+}{\partial z} = &-i\left[\operatorname{Re} H_{N}, \rho_{+}\right]+\frac{1}{2}\left[\operatorname{Im} H_{N}, \rho_{-}\right]-\frac{1}{2}\left\{\operatorname{Re} \Gamma_{N}, \rho_{+}\right\}-\frac{i}{4}\left\{\operatorname{Im} \Gamma_{N}, \rho_{-}\right\}\nonumber\\
&-\frac{i}{2} \sum\left(\operatorname{Im} \tilde{\Gamma}_{N}^{\alpha}\right) 2\,\tilde{\omega}_{\alpha \beta}(T) Y_{\Delta_\beta}\frac{s}{T^3}\, f_\nu\,(1-f_\nu)-S^{eq},\\
H\, \frac{\partial \rho_-}{\partial z} = &
2\left[\operatorname{Im} H_{N}, \rho_{+}\right]-i\left[\operatorname{Re} H_{N}, \rho_{-}\right]-i\left\{\operatorname{Im} \Gamma_{N}, \rho_{+}\right\}-\frac{1}{2}\left\{\operatorname{Re} \Gamma_{N}, \rho_{-}\right\}\nonumber\\
&-\sum\left(\operatorname{Re} \tilde{\Gamma}_{N}^{\alpha}\right)\,
2\,\tilde{\omega}_{\alpha \beta}(T) Y_{\Delta_\beta}\frac{s}{T^3}\, f_\nu\,(1-f_\nu),
\end{align}\label{kin_eq_mom}\end{subequations}
where $H$ is again the Hubble rate, $s$ is the entropy density, and $S^{eq} = \frac{\partial \rho^{eq}_N}{\partial z}$. Let us stress that both $\rho_+$ and $\rho_-$ depend on the momentum $y$. 

In order to solve this set of integro-differential equations, we introduce a grid in the momentum $y$ and consider a fixed number of modes (denoted as $N_{modes}$). The integrals in the r.h.s.\ of~\eqref{ke1} can be computed using a simple quadrature. 
Practically we have verified that the trapezoidal rule with a reasonably chosen momentum grid is sufficient for our purposes. A rather natural choice of the momentum grid can be the following. For a given number of modes $N_{modes}$ we define $y_i$ by solving
\begin{equation}
    \int\limits_{y_i}^{y_{i+1}} \frac{d^3y}{(2\pi)^3} \frac{1}{\exp(y)+1} / 
    \int\limits_{y_1}^{y_{max}} \frac{d^3y}{(2\pi)^3} \frac{1}{\exp(y)+1} = 
    1/ (N_{modes} - 1),
\end{equation}
where $y_1$ is the lowest accessible momentum (in our case it is $0.1$) and $y_{max} = 20$.  A greed defined this way can be improved by manually adding some low momenta modes ($y = 0.3,\, 0.5,\, \ldots$). 

\begin{figure}[h!]
    \centering
    \includegraphics[width=0.8\textwidth]{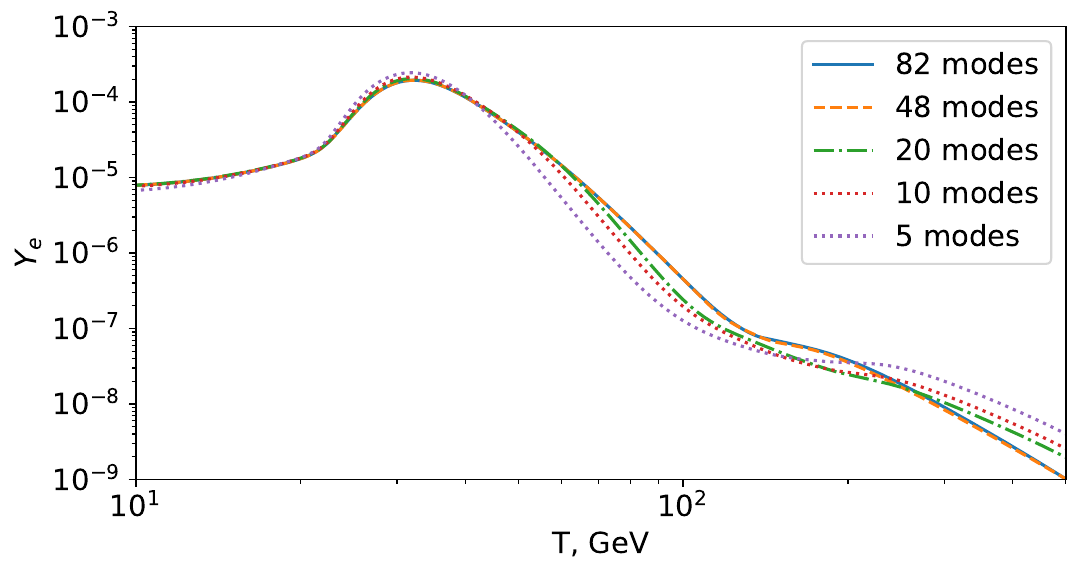}
    \caption{Comparison of the solutions ($Y_e$ is shown) with different numbers of the momentum modes. The model parameters are specified in eq.~\eqref{example_parameters}. As one can see, the system approaches an accurate solution when the number of the momentum modes is large enough.}
    \label{fig:momentum_modes}
\end{figure}

We have implemented the set of $8\times N_{modes} + 3$ equations in \texttt{Julia} language~\cite{Julia-2017} and solve them numerically by means of \texttt{Differentialequations.jl} package~\cite{rackauckas2017differentialequations} (again using LSODA solver~\cite{LSODE}). We solved eqs.~\eqref{kin_eq_mom} varying $N_{modes}$ from $5$ to $82$. 
This allowed us to verify that for the fine enough grids, the solution no longer depends on $N_{modes}$. This is illustrated in figure~\ref{fig:momentum_modes}.

\begin{figure}[h]
    \centering
    \includegraphics[width=0.8\textwidth]{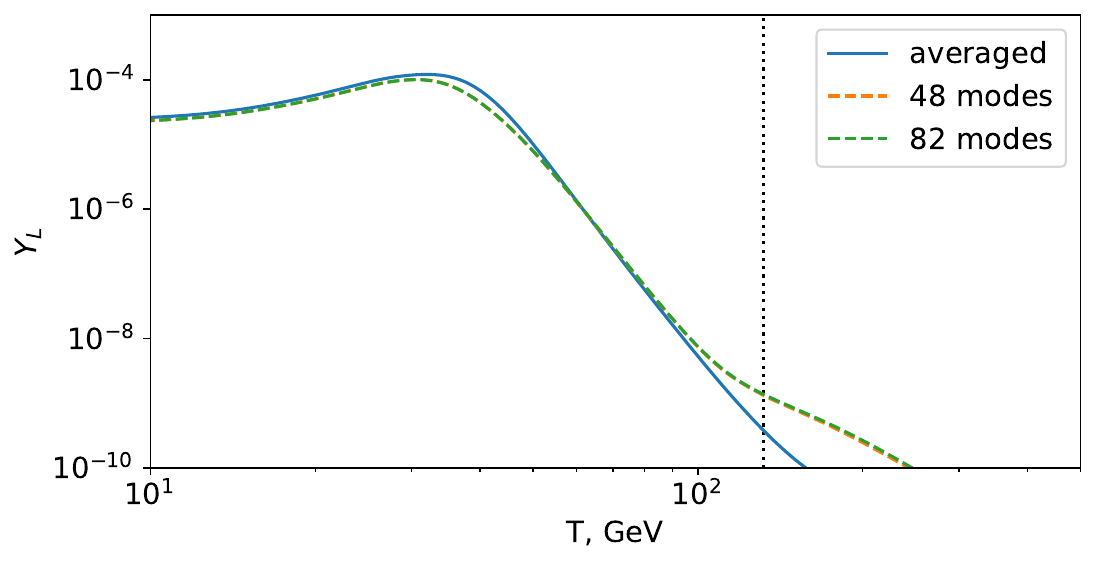}
    \caption{The total lepton asymmetry as a function of temperature. At low temperatures, the average solution (solid blue line) is very close to the accurate momentum dependent solution with 48 (orange dashed line) and 82 (green dashed line) momentum modes. The black dotted vertical line shows the temperature of the sphaleron freeze-out.  The model parameters are specified in eq.~\eqref{example_parameters}.}
    \label{fig:momentum_dependent}
\end{figure}

Of course, solving $8\times N_{modes} + 3$ with $N_{modes} = 48$ is numerically very demanding. A systematic study analogous one in section~\ref{sec:results} is thus prohibitively expensive in terms of the computer resources.
We can, however, directly compare the results of the averaged and accurate equations for selected parameter sets. We have identified several such sets yielding large LTA and performed the comparison. We have found that for these sets the averaged result agree with the accurate ones within a factor of $\sim1.3$. One particular example of this comparison is shown in figure~\ref{fig:momentum_dependent}.
For the parameter sets resulting in much smaller values of LTA we found a factor of two agreement between the averaged and accurate equations.

One can notice that while the values of LTA is almost the same, the asymmetries at the sphaleron freeze-out differ in two approaches. As a result, the value of the BAU computed using the averaged equations differs from the accurate one. This has two consequences. First, a careful analysis of BAU by means of momentum-dependent equations will be very interesting. We leave this for future work. Secondly, even though the averaged equations describe LTA generation with $\sim 30 \%$ accuracy, the BAU values may change, and therefore the green bands in figures~\ref{fig:LTA} and~\ref{fig:LTA_IH} also change. This, however, does not affect our conclusions about DM production.


\section{Chiral asymmetry and magnetic fields} 
\label{sec:chiral_asymmetriy_and_magnetic_fields}

Up to date,  all quantitative studies of the low-scale leptogenesis were based on kinetic equations (\ref{kin_eq}). 
These equations seem to account for all relevant processes except those related to the Abelian part of the anomaly (second term in the equation below) in the leptonic  $j^\mu_L$ and baryonic $j^\mu_B$ currents,
\begin{equation}
\label{anomaly}
\partial_\mu j^\mu_L=\partial_\mu j^\mu_B=\frac{n_f}{32\pi^2}\epsilon^{\mu\nu\rho\sigma}\left(\frac{1}{2}g^2 F^a_{\mu\nu} F^a_{\rho\sigma}+\frac{1}{2}g'^2 F_{\mu\nu} F_{\rho\sigma}\right)~.
\end{equation}
Here $n_f=3$ is the number of fermionic generations, $F^a_{\mu\nu}$ and $F_{\mu\nu}$ are the $SU(2)$ and $U(1)$ gauge field strengths respectively.  The Abelian field contribution to the divergence of the currents does not lead to the irreversible change of the baryon and lepton numbers, since for $U(1)$ gauge fields the Chern-Simons (CS) charge for vacuum configurations is zero. Still, if the U(1) field is massless (as is the case for the hyper-charge field in the symmetric phase or electromagnetic field in the Higgs phase), the non-vacuum configurations with non-zero (hyper) magnetic field may carry CS number and therefore the transfer of baryon or lepton numbers into CS ``condensate'' of (hyper) magnetic fields may take place. And, indeed, if there were primordial helical hyper-magnetic fields, they could be converted into baryon asymmetry \cite{ Giovannini:1997eg,Kamada:2018tcs}. Or, the large fermionic asymmetries can induce instabilities in the gauge sector and lead to the generation of magnetic fields \cite{Joyce:1997uy, Boyarsky:2011uy}.

We will assume that there are no primordial (hyper) magnetic fields. If true, the effects associated with the Abelian part of the anomaly equation can be neglected for analysis of the baryogenesis.  Indeed,  in the symmetric phase of the SM the rate of the $SU(2)$ anomalous processes is much higher than that related to the $U(1)$ group. Also,  the generation of magnetic fields is a non-linear effect \cite{Giovannini:1997eg,Boyarsky:2011uy}  which requires large fermionic asymmetries. However, they are very small at the sphaleron freeze-out (we know this empirically from the measured value of the baryon asymmetry) and the system stays in the linear regime.

The generation of (large) lepton asymmetry can only occur in the Higgs phase of the SM, below the sphaleron decoupling. Here the massless gauge degree of freedom which can give rise to long-ranged field corresponds to the electromagnetic field $A_\mu$, which is a mixture of the $SU(2)$ and $U(1)$ hypercharge fields. We will denote the corresponding magnetic and electric field strengths by $\vec{B}$ and $\vec{E}$. The right-hand side of eq.~\eqref{anomaly} {\em does not contain}  a product  $\vec{B}\vec{E}\propto F_{EM}{\tilde F}_{EM}$ meaning that the baryon and lepton numbers cannot be converted into the CS condensate of the electromagnetic field. The combination $F_{EM}{\tilde F}_{EM}$ still appears as an anomaly in the chiral current of leptons,
\begin{equation}
\label{chiral}
\partial_\mu j^\mu_{5\alpha}= m_\alpha{\bar \Psi}_\alpha\Psi_\alpha + \frac{1}{32\pi^2}\epsilon^{\mu\nu\rho\sigma}\frac{1}{2}e^2 F^{EM}_{\mu\nu} F^{EM}_{\rho\sigma}~,
\end{equation}
where $m_\alpha$ is the mass of a lepton of generation $\alpha$ and $e$ is the electric charge. The reactions which change the lepton chiralities can occur due to perturbative spin-flip (the first term in (\ref{chiral}))  or due to the processes in which the CS number of the electromagnetic field is changed (the second term in (\ref{chiral})). The rate of the first type of reactions at temperatures higher than the mass of the corresponding lepton is at least of the order of $\Gamma_{flip} \sim\frac{\alpha^2}{4\pi}\frac{m_l^2}{T}$.\footnote{It has been shown in the recent works~\cite{Boyarsky:2020cyk,Boyarsky:2020ani} that this rate is in fact much larger.} These are  just electromagnetic reactions: the Compton process $l\gamma\to l\gamma$ and annihilation  $l^+l^-\to\gamma\gamma$ \footnote{Yet another contribution is the Higgs decays, $h\to l^+l^-$, which freezes out at temperature $\simeq 8$ GeV for electron flavour \cite{Boyarsky:2011uy} must be considered as well.}.

The rate of the reactions of the second type is known to be non-zero in the presence of external magnetic fields. An evaluation based on magnetohydrodynamics (MHD) reads \cite{Joyce:1997uy,Giovannini:1997eg} $\Gamma_{anom}=\frac{12 \alpha^2}{\pi^2\sigma} \frac{B^2}{T^2}$, where $\sigma\propto T/\alpha$ is the conductivity of the plasma. This estimate accounts only for the fluctuations of the electromagnetic field on the scales larger than the mean free path of the particles in the plasma. The recent lattice simulations \cite{Figueroa:2017hun,Figueroa:2019jsi} incorporating the short-scale fluctuations indicated that the actual rate has the same parametric dependence on the magnetic field and the fine-structure constant $\alpha$, but is larger by a factor of 10. It is not excluded that the non-perturbative processes can occur even in the absence of the magnetic fields (see the discussion in \cite{Figueroa:2017hun}) with the rate of the order of $\alpha^6 T$.  These processes are of a similar nature to the weak SU(2) sphaleron processes\footnote{The importance of accounting for fluctuations of Abelian U(1) field in anomalous processes was also stressed in \cite{Long:2013tha}.}.

In the processes of the first type, the fermionic chirality is lost forever, while in the processes of the second type, it is transferred into Chern-Simons number of the $U(1)$ field (it can be called CS condensate). If the chirality flip rate is inferior to the one due to anomaly, the entire chiral charge is converted into the CS condensate. It is the electron flavour that is the most important as the perturbative chirality flip rate for it is much smaller than for the other flavours. The evolution of the chiral charge and of the magnetic field is quite peculiar, it has been studied in \cite{Joyce:1997uy,Boyarsky:2011uy}. Basically, the system enters the steady-state {\em non-linear}  evolution in which the chiral charge and the CS condensate of electromagnetic fields change much slower than one would expect from the linear analysis and from the rates of different reactions: the anomaly pumps the chiral charge into the CS condensate, the CS condensate decreases due to plasma conductivity and releases back the chiral charge into the plasma.  These two processes nearly compensate each other, leading to approximate conservation of the chiral charge. 

The phenomena discussed above may result in modification of the evolution of leptonic asymmetries, which we described up to now by equations \eqref{kin_eq} below the sphaleron freeze-out.  The detailed study of this problem would require three-dimensional magneto-hydro-dynamical simulations including helical and non-helical magnetic fields, unified with analysis of the thermal fluctuations of electromagnetic fields at smaller scales. We leave this for future work.  Instead, we will discuss at the quantitative level what kind of physical effects one can expect and how they can modify the computation of LTA.

The important point is that HNLs interact directly {\em only with the left-handed leptons}. Therefore, their CP-violating interactions produce not only the leptonic asymmetry, but also the asymmetry in the chiral charges, and, in particular, in the chiral lepton densities.  At temperatures where the production of the lepton asymmetries is the most efficient, $T\sim 20$ GeV,  the created chiral charge may be large enough to lead to the generation of the helical magnetic field, and the anomalous rate $\Gamma_{anom}$  exceeds the perturbative chirality flip rate $\Gamma_{flip}$ for electron, but $\Gamma_{anom} \lesssim \Gamma_{flip}$ for muon and tau leptons. This means that the chiral asymmetry in muons and tau flavours is destroyed, but that sitting in the electron flavour is transferred to the CS condensate.  If it is sufficiently large, the system enters into the non-linear regime as above with effectively conserved chiral asymmetry in the electronic flavour. The processes with HNLs will redistribute this asymmetry among the other flavours forming a  long-living configuration carrying a net {\em lepton}  number density and magnetic helicity.
 In the most optimistic scenario, the maximal asymmetry in the electronic flavour, attained during the time evolution that does not account for magnetic fields, will survive until the late times and amplify the sterile neutrino DM production.

Having these physics considerations in mind it is possible to write ``phenomenological'' equations accounting for CS condensate. To this end let us introduce as usual the chemical potentials $\mu_i$ for slowly varying  leptonic numbers $L_i$, for electric charge $Q$ and baryon number $B$:  $\mu_q$ and $\mu_B$, and, to account for CS condensate, the chemical potential for the number density $R_0$ of the right-handed electrons, $\mu_R$ (when the processes with chirality flip are in thermal equilibrium and anomaly is absent $\mu_R=0$). The standard procedure allows to express $\mu_i$ and $\mu_R$ via $L_i$ and $R_0$ (we put $Q=0$, and also $B=0$ since the baryon asymmetry is too small to lead to any effects). The result is
\begin{align}
\mu_1&=\frac{6}{271\,T^2}\left(143 L_1 +  10 L_2  +   10 L_3 +128 R_0\right),\\
\mu_2&=\frac{2}{271\,T^2}\left(30   L_1 + 311 L_2  +   40 L_3 -   30 R_0\right),\\
\mu_3&=\frac{2}{271\,T^2}\left(30   L_1 + 40   L_2  + 311 L_3 -   30 R_0\right),\\
\mu_R&=\frac{12}{271\,T^2}\left(64 L_1 -  5    L_2  -      5 L_3 + 207 R_0\right).
\label{muR}
\end{align}
The expressions for $\mu_i$ should be used in equations (\ref{kin_eq}) written in previous chapters of this work.\footnote{In this section, we consider temperatures below $T_{sph}$ and do not distinguish lepton asymmetries $L_\alpha$ and $L_\alpha-B/3$ owing to the fact $B/s\sim 10^{-10}$ whereas the lepton asymmetries interesting from the DM production perspective are $5$ orders of magnitude larger.}  To account phenomenologically for magnetic fields, we should add an equation for $R_0$. In the absence of the CS condensate and of anomalous reactions, the equation for $R_0$ has the form
\begin{equation}
\frac{\partial R_0}{\partial t} = -\Gamma_R \mu_R~,
\label{stand}
\end{equation}
where $\Gamma_R$ is the perturbative chirality flip rate. This equation will drive $\mu_R$ to zero, and  one would find that the chiral asymmetries in the electronic flavour are washed out at temperature $T_{eq} \simeq 8.5\times10^4$ GeV (the most accurate discussion of the equilibration of right-handed electrons is contained in \cite{Bodeker:2019ajh}). If only the perturbative reactions were present, the asymmetries below this temperature would be exponentially suppressed by the factor $\exp(-\Gamma_R dt)$. However, the studies of \cite{Joyce:1997uy,Giovannini:1997eg,Boyarsky:2011uy} demonstrated that when the magnetic field is generated due to the instability induced by the chiral anomaly and the presence of the chiral charge gets sufficiently large (i.e. when $\Gamma_{anom} \gtrsim \Gamma_{flip}$), the system enters in a steady-state regime. In this regime, the chemical potential and the magnetic energy are changing much slower in comparison with the exponential perturbative behaviour. 
In addition, the non-perturbative effects related to the $U(1)$ fields are only essential for the electron family and do not touch the $\mu$ and $\tau$ flavours. This enhances the CP-violating effects removing cancellations between leptonic generations present in the system of kinetic equations~\eqref{kin_eq}.
It is these effects that may provide a possibility for the leptonic asymmetries to survive below $T_{eq}$.

Now, if the CS condensate and anomalous reactions are present, $\mu_R$ will be driven to some fraction $\kappa$ of the asymmetry in $L_1$, resulting in the change of eq. (\ref{stand}) to
\begin{equation}
\frac{\partial R_0}{\partial t} = -\Gamma_C\left(\mu_R-\kappa\mu_1\right)~,
\label{anom}
\end{equation}
where $\Gamma_C$ is the effective rate that accounts for non-linear dynamics of the condensate and the chiral charge. When the combination $\left(\mu_R-\kappa\mu_1\right)$ approaches zero, the system should enter the steady state solution with non-zero value of $R_0$. This is achieved if $\Gamma_C$ also goes to zero when $\left(\mu_R-\kappa\mu_1\right)\to 0$. Assuming $\Gamma_C = \gamma_C \, \left(\mu_R-\kappa\mu_1\right)^2$ we arrive at an effective equation
\begin{equation}
\frac{\partial R_0}{\partial t} = -\gamma_C\left(\mu_R-\kappa\mu_1\right)^3~,
\label{eff}
\end{equation}
where $\gamma_C$ does not depend on the chemical potentials.
In the steady-state solution derived in \cite{Joyce:1997uy,Giovannini:1997eg, Boyarsky:2011uy} the ${\cal O}(1)$ fraction of the initial asymmetry in right-handed electrons goes into Chern-Simons condensate of the Abelian gauge field. The remaining right-handed electrons can be transferred into left-handed electrons by the chirality flip reactions, and subsequently left-handed electrons into neutrinos by the equilibrium weak interactions. We expect, therefore, that in the steady-state the chemical potentials $\mu_R$ and $\mu_1$ are of the same order, leading to the conclusion that $\kappa$ should be of the order of one. Now, in the steady-state the density of the right-handed electrons is non-zero, meaning that the eq.~\eqref{anom}considered together with \eqref{kin_eq} should have solutions with $R_0=const \neq 0$. We verified numerically that if $\Gamma_C \neq 0$ at $\mu_R - \kappa\mu_1 = 0 $ this is not the case. From here, it is natural to expect that $\Gamma_C$ should be some power of $\mu_R - \kappa\mu_1$. The first power does not work as then the equation is unstable close to $\mu_R - \kappa\mu_1 = 0 $. Therefore, we chose the second power with the sign that makes the point $\mu_R - \kappa\mu_1 = 0$ to be an attractor. The choice of $\gamma_C$ is motivated by the following arguments. In the steady-state the chemical potential and the magnetic energy are changing as \cite{Joyce:1997uy,Giovannini:1997eg,Boyarsky:2011uy} $\mu_R/T \propto B^2/T^4 \propto (T/T_{\rm inst})^{\frac{1}{2}}$, where $T_{\rm inst}$ is the temperature at which the magnetic instability develops, $T_{\rm inst}\simeq 8 M_0 \left(\mu_R/T\right)^2(T/\sigma)$. The temperature behaviour of the chemical potential $\mu_R$ is reproduced if $\gamma_C \propto H$, where $H$ is the Hubble expansion rate. 

\begin{figure}[h]
    \centering
    \includegraphics[width=0.6\textwidth]{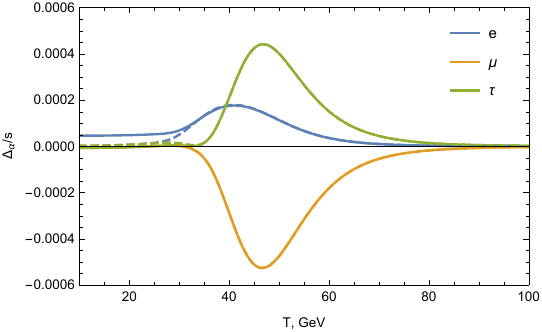}
    \includegraphics[width=0.6\textwidth]{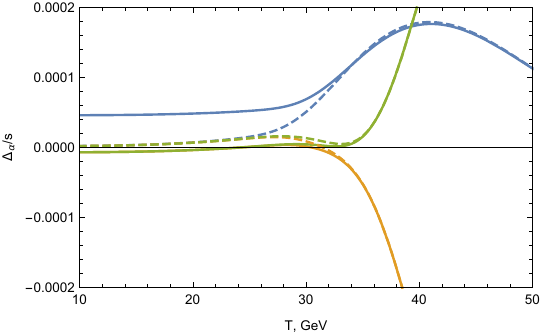}
    \caption{Evolution of the lepton asymmetries in the presence of chiral condensate (solid curves). For comparison, we show the evolution described by the equations \eqref{kin_eq} without \eqref{eff} (dashed curves). In the presence of chiral condensate the electron asymmetry freezes at much larger value. The lower panel shows the zoomed in region 
    $T < 50$~GeV. Here $\kappa = 0.1$, while $\gamma_C = 10^4 \times H$ at temperatures above $\simeq 40$~GeV and quickly drops at lower temperatures. }
    \label{fig:magn_evol}
\end{figure}

The system of eqs.~\eqref{kin_eq} and \eqref{eff} contains two new parameters, $\gamma_C$ and $\kappa$, the estimate of which goes beyond the scope of this paper. Here, we simply considered the evolution of the system in different cases. A particular example with $\kappa = 0.1$ is shown in Fig. \ref{fig:magn_evol}. 
In this example, the rate $\gamma_C$ is much larger than Hubble at high temperatures and then quickly drops at the temperature where the electron asymmetry is maximal. This choice of parameters is entirely ad hoc and just serves the purpose of showing that they can be picked in a way that leads to asymmetry enhancement.
The number density of right-handed electrons freezes at its maximum value. The resulting lepton asymmetry still ``feels'' the effects of HNLs that freeze a bit later. This leads to the partial redistribution of electron asymmetry in agreement with \eqref{muR}. Still, as one can see from figure~\ref{fig:magn_evol}, the resulting electron asymmetry is much larger than asymmetries in the other lepton flavours, whereas if the dynamics of chiral condensate is not accounted for, all three asymmetries are the same (cf. the dashed line in figure~\ref{fig:magn_evol}).

Let us stress that the helical magnetic fields hide the electron asymmetry from washout at the moment when the asymmetries are generated, at $T\sim 40$ GeV for the choice of the parameters leading to figure~\ref{fig:magn_evol}. This results, in turn, in the change of the balance between the asymmetries in different flavours and leads to the larger values of the approximately conserved charge $L_+$ defined in eq.~\eqref{conserved_combination}, see figure~\ref{fig:magn_evol}. The charge $L_+$ survives to the low temperatures relevant for DM production. So, the magnetic fields are only relevant in the domain of temperatures where the generation of the asymmetry is the most efficient, $T\sim 40$ GeV, and their role is to put more asymmetries into $L_+$. The helical magnetic fields are not converted back into the lepton asymmetries.

As we have already discussed, the most optimistic case corresponds to the pattern in which the asymmetry in electron flavour reaches its maximum at some temperature and then freezes around this value. By varying the coefficients $\gamma_C$ and $\kappa$  we were able to see that this type of behaviour can always be achieved. So, for the scan of the parameters in this most optimistic case, we were solving the original equations (\ref{kin_eq}) and saving the maximal asymmetry in the electron flavour attained during the time evolution. This procedure is discussed in greater detail in the next section.

\section{Maximal electron asymmetry} 
\label{sec:maximal_electron_asymmetry}
In the previous section, we have seen that the processes related to the Abelian part of the anomaly can significantly alternate dynamics of asymmetry production if large electron asymmetry is generated at some temperature.

\begin{figure}[h!]
   \centering
  \includegraphics[width=0.6\textwidth]{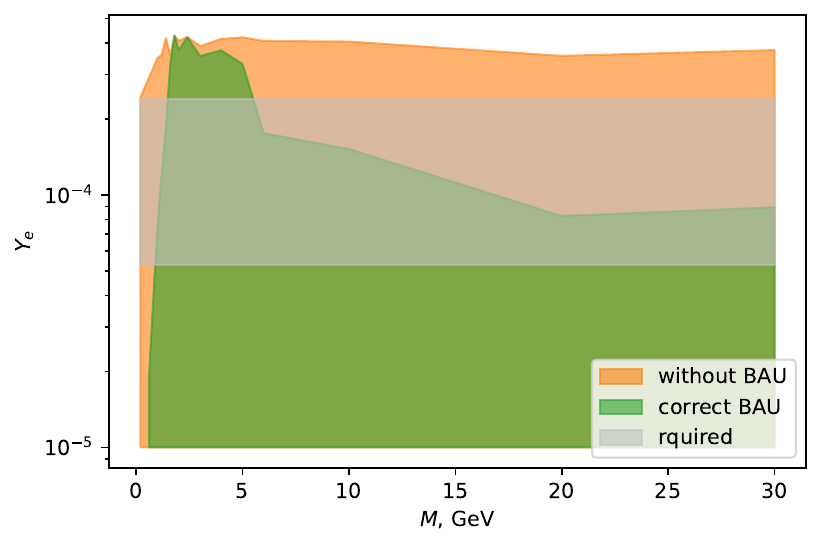}
  \includegraphics[width=0.6\textwidth]{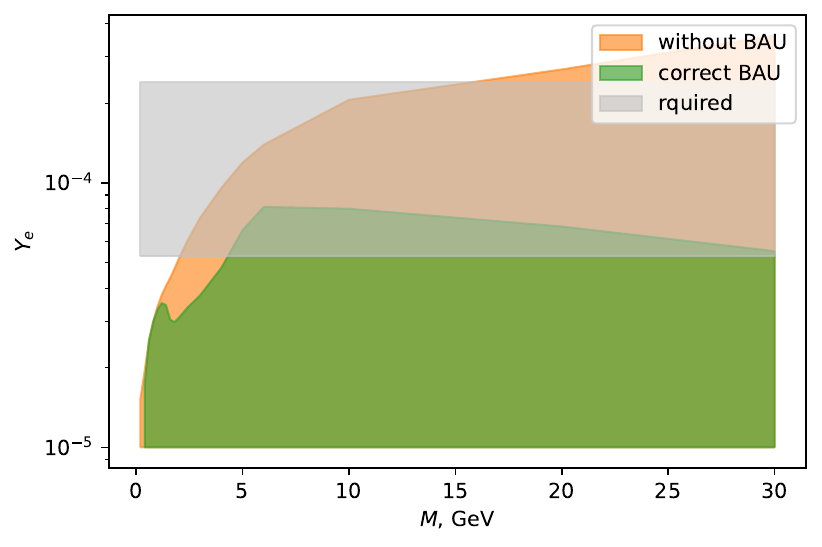}
  \caption{Maximal electron asymmetry as a function of M. \emph{Upper panel:} normal hierarchy. \emph{Lower panel:} inverted hierarchy. The grey shaded region indicates the electron asymmetry which is needed for $N_1$ to compose $100\%$ of DM. The span over $\Delta_e$ is due to different assumptions about DM mixings. Thickness of the orange and green curves indicates  the possible uncertainty of the averaging procedure. }
  \label{fig:maxE}
\end{figure}

It is premature to perform a parameter scan using eqs.~\eqref{kin_eq} \emph{plus}~\eqref{eff} since we don't know the exact values (and temperature dependence) of the coefficients $\gamma_C$ and $\kappa$. Still, it is important to clarify the potential significance of the effects related to the anomaly. 
To this end, we have asked a different question: what is the maximal value of the electron asymmetry which can be reached in the system described by eqs.~\eqref{kin_eq}. This maximal asymmetry can be partially conserved as in the example of figure~\ref{fig:magn_evol}.
Let us stress that the generation of maximal asymmetry does not guarantee that the final asymmetry will be the largest. 
In fact, there are two processes that are important: \emph{\emph{(i)}} generation of the chiral asymmetry and \emph{(ii)} its wash-out due to interactions with the HNLs. Therefore, the parameter leading to the largest value of the maximal electron asymmetry does not necessarily lead to the largest value of the final asymmetry, since redistribution between the other flavours can be more efficient as well. 
Nevertheless, a study of the maximal possible electron asymmetry is the first step in this direction.  The results of the scan of the parameter space are presented in figure~\ref{fig:maxE}.  In this figure, we show the maximal value of the asymmetry in the electron flavour as a function of the mass of the HNLs. As one can see, the maximal asymmetry can be of the correct magnitude for HNLs with $M>2$~GeV. Even though this fact doesn't guarantee that the $\nu$MSM can account for  $100\%$ of DM without fine-tunings of ref.~\cite{Canetti:2012kh}, it indicates the importance of the non-perturbative effects which have not been accounted for so far.

\section{Possible uncertainties of kinetic equations} 
\label{sec:possible_uncertainties_of_kinetic_equations}
We have already mentioned in section~\ref{sec:generation_of_asymmetry} that our analysis contains the inherent uncertainty related to the integration of the kinetic equations over momentum. We have also stressed that the non-perturbative effects related to Abelian chiral anomaly are not accounted for in the kinetic equations. In this section, we discuss the other possible sources of uncertainties.

\begin{figure}[h]
    \centering
    \includegraphics[width=0.48\textwidth]{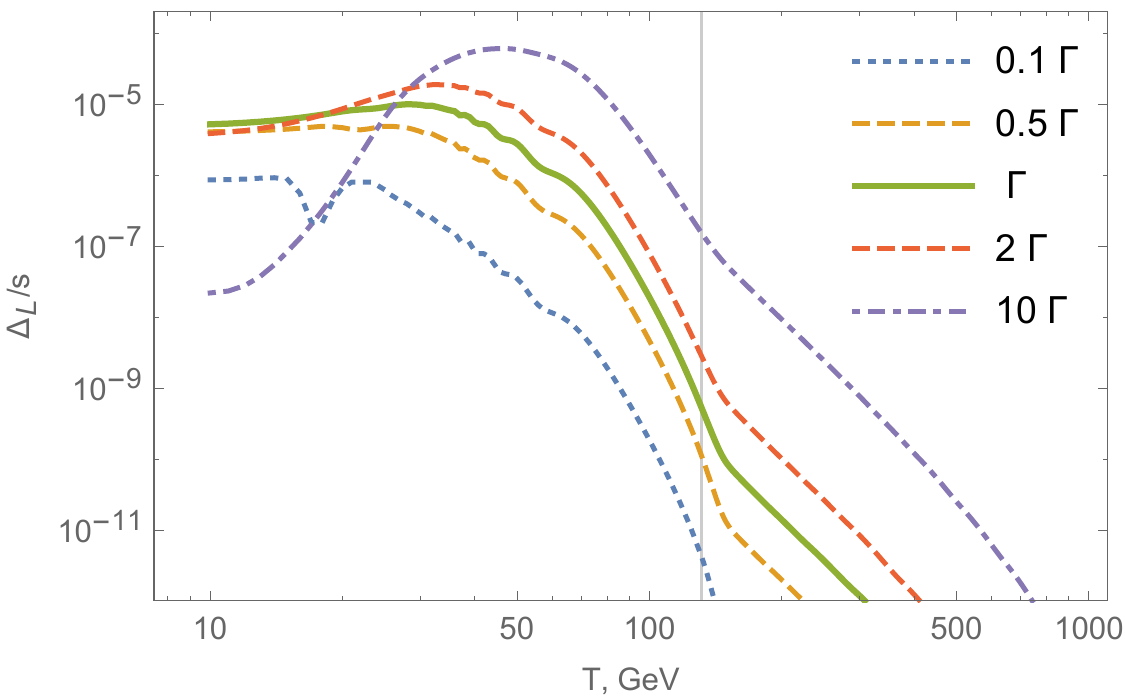}
    \includegraphics[width=0.48\textwidth]{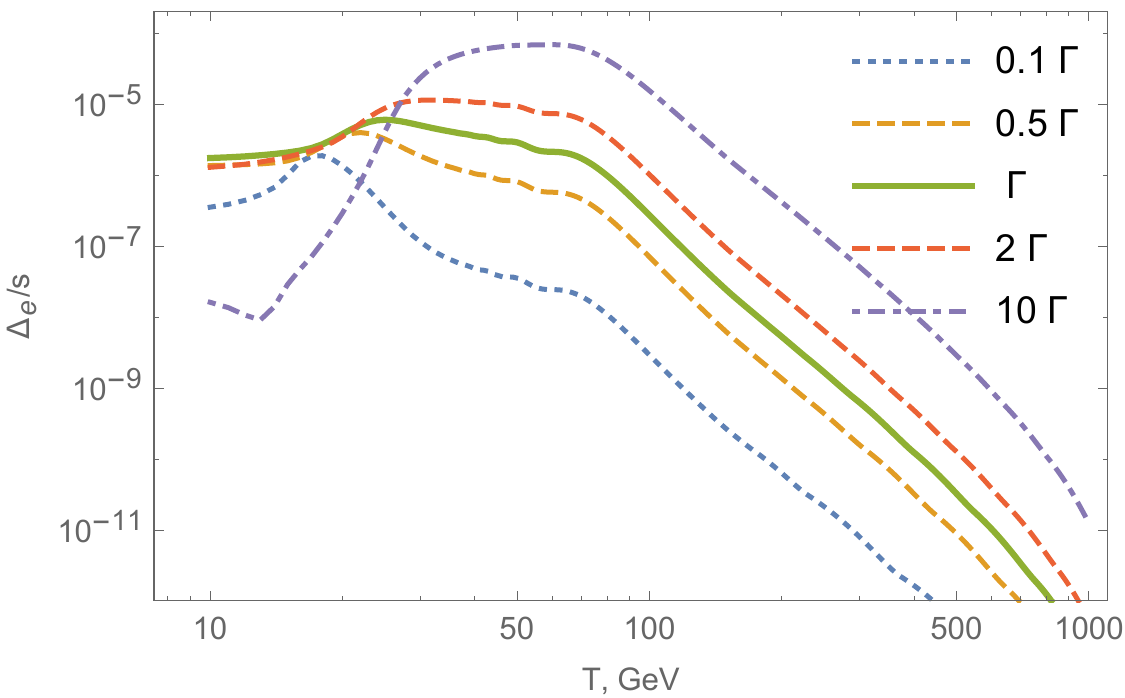}
    \caption{Evolution of the total lepton asymmetry (left panel) and the electron asymmetry (right panel) computed using the rates multiplied by a constant factor $\Gamma \to \kappa \cdot \Gamma$, where $\Gamma$ denotes all rates $\Gamma_{\nu_\alpha}, \; \tilde{\Gamma}_{\nu_\alpha}, \; \Gamma_N,\; \tilde{\Gamma}_{N}^\alpha$. The solid green line corresponds to the original rates. The vertical grey line in the left plot indicates the sphaleron freeze-out temperature $T_{sph}\simeq 131.7$~GeV. The parameters are fixed to the values $M=1$~GeV, $\Delta M = 1.26\times10^{-11}$~GeV, $\imw = -0.14$, $\rew = 0.52 \pi$, $\delta = 1.30 \pi$, and $\eta = 1.02 \pi$.}
    \label{fig:varying_rates}
\end{figure}

First, let us comment on the role of the rates entering kinetic equations~\eqref{kin_eq}. In the broken phase these rates can be split into ``direct'' (proportional to $T$) and ``indirect'' (proportional to the Higgs vev $v_0$) parts~\cite{Ghiglieri:2016xye}. The ``indirect'' rates depend on the thermal neutrino interaction rates.  Recently, one of these rates has been computed at next-to-leading order~\cite{Jackson:2019tnr}. It has been found that the NLO rate is $15\ldots 40 \%$ smaller than the leading order one.  
Even though it has been shown in ref.~\cite{Jackson:2019tnr} that the resulting lepton asymmetry is affected only at $1\%$ level, it is still reasonable to question the sensitivity of LTA to variation of the other rates.
To partially clarify this issue we have computed LTA multiplying the rates $\Gamma_{\nu_\alpha}, \; \tilde{\Gamma}_{\nu_\alpha}, \; \Gamma_N,\; \tilde{\Gamma}_{N}^\alpha$ entering kinetic equations~\eqref{kin_eq}, by a constant factor.
Result of this procedure is shown in figure~\ref{fig:varying_rates}.
The resulting LTA is surprisingly stable against the factor of $2$ variations of the rates. Note, however, that the value of the BAU is sensitive to these variations. Namely, if the rates are larger or smaller, one will need to change the other parameters of the theory in order to obtain the correct amount of BAU, this, in turn, will change the resulting lepton asymmetry. 
Further, figure~\ref{fig:varying_rates} shows that larger rate variations completely change the dynamics.
Also, as one can see from the right panel, the maximal value of the electron asymmetry is more sensitive to the rates. Therefore, further refinement of the rates is an important task. 

A more serious concern is related to the effective Hamiltonian.
Indeed, the derivation of equations \eqref{kin_eq} in refs.~\cite{Eijima:2017anv,Eijima:2018qke}
was based on the separation of time scales. Namely, all quantities proportional to Majorana mass difference of HNLs $\Delta M$ or to Yukawas  $F_{\alpha I} v_0$ were treated as small compared to  combinations of energies of neutrinos and HNLs,
such as $E_N+E_\nu$ or $E_N-E_\nu$:
\begin{equation}
    \Delta M, \; F_{\alpha I} v_0 \quad \ll \quad  E_N+E_\nu,\; E_N-E_\nu.
    \label{separation_of_scales}
\end{equation}
This allowed us to integrate out the fast processes and derive equations \eqref{kin_eq} in terms of slowly varying quantities. 
In the mass region of interest, $M \lesssim 35~$GeV, assumption \eqref{separation_of_scales} is  justified at temperatures above the sphaleron freeze-out, $T_{sph}\simeq 131.7$~GeV.
In this regime, thermal corrections to active neutrino energy are proportional to $T$ and large~\cite{Weldon:1982bn}. However, \eqref{separation_of_scales} might not be valid at temperatures around a few tens of GeV. The reason is that the thermal correction to active neutrino at some point changes its sign, so the energy levels of HNLs and active neutrinos do cross. 
This effect has not been accounted for in our rates in the Higgs phase. 

\begin{figure}[h]
    \centering
    \includegraphics[width=0.48\textwidth]{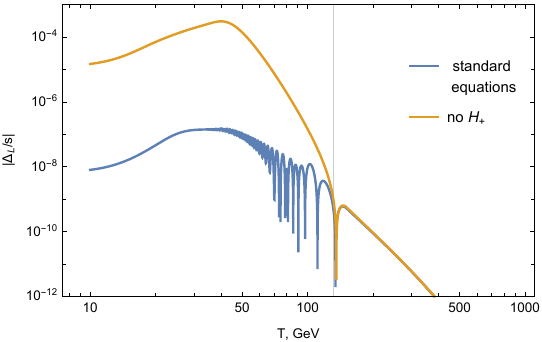}
    \includegraphics[width=0.48\textwidth]{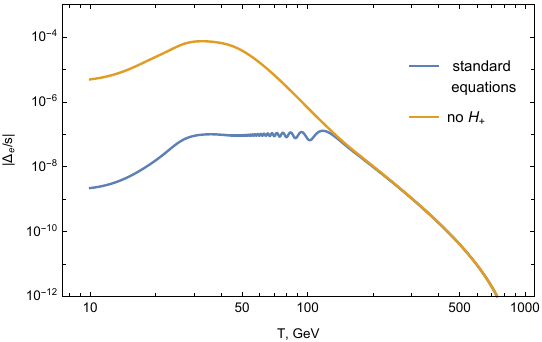}
    \caption{Evolution of the total lepton asymmetry (left panel) and the electron asymmetry (right panel) 
    computed with (blue line) and without (orange line) $H_+$ part of the effective Hamiltonian.
    The parameters are fixed to the values $M=1$~GeV, $\Delta M = 1\times10^{-12}$~GeV, $\imw = \log(3)$, $\rew = 13/16 \pi$, $\delta = 29/16 \pi$, and $\eta = 22/16 \pi$.}
    \label{fig:no_hplus}
\end{figure}

While the LTA calculations are rather robust to changes of the rates, the role of the effective Hamiltonian $H_N$ is far more drastic. The effective Hamiltonian can be decomposed into fermion number conserving and violating parts: $H_N = H_+ + H_-$ (following notations of refs.~\cite{Eijima:2017anv,Eijima:2018qke}). For an illustration we  switched off the $H_+$ part (see the precise expressions in~\cite{Eijima:2017anv,Eijima:2018qke}) completely and solved the kinetic equations, see figure~\ref{fig:no_hplus}.\footnote{Let us note in passing that computing BAU for the  parameter set used in figure~\ref{fig:no_hplus} requires one to use the approach of ref.~\cite{Eijima:2017cxr}, since the total lepton asymmetry changes the sign right before the sphaleron freeze-out. This approach allows for accurate tracking of BAU without enlarging the set of equations and making it stiffer. To this end one needs to solve a separate kinetic equation for sphalerons using the lepton asymmetries as an input.}
It shows that the production of the asymmetry is greatly enhanced if $H_+=0$. The reason is that $H_-\sim (E_N - E_\nu)$
crosses zero at some temperature around $20-30$~GeV, precisely where the rates peak. Around the zero of $H_-$ the physical mass difference is determined only by the Majorana part, and can be small enough to ensure the resonant amplification of the asymmetry production. Similar observations about the importance of the thermal mass corrections have been made in ref.~\cite{Ghiglieri:2018wbs}.

Figure~\ref{fig:no_hplus} illustrates the importance of the effective Hamiltonian. Both LTA and maximal electron asymmetry can vary by many orders of magnitude depending on the behaviour of $H_I$ around the point where $E_N=E_\nu$. Therefore a dedicated derivation of the kinetic equations accounting for the level crossing is very desirable. We leave this for future work.


\section{Discussion} 
\label{sec:discussion}
In this work, we have studied the freeze-in and freeze-out generation of the late time asymmetry in the $\nu$MSM. More specifically, we have systematically investigated the production of the asymmetry which takes place before decays of HNLs. Using the momentum averaged kinetic equations based on (\ref{kin_eq}) with the state-of-the-art rates taken from~\cite{Ghiglieri:2018wbs} we have found that large lepton asymmetry can be generated at temperatures above a few GeV. The generation of the asymmetry takes place during the freeze-in of the HNLs. This mechanism differs from the previously considered: refs.~\cite{Canetti:2012vf,Canetti:2012kh} were considering the freeze-out and decays, whereas ref.~\cite{Ghiglieri:2020ulj} considered the decays only. The freeze-in mechanism is operative due to the presence of both fermion number violating and conserving rates~\cite{Eijima:2017anv,Ghiglieri:2017gjz}. Even though the generated asymmetry could be quite large, the subsequent entropy injection from HNL decays reduces the created DM abundance. 
Using the benchmarks from ref.~\cite{Ghiglieri:2015jua} we found that the asymmetry generated during freeze-in could eventually be responsible for $\sim 50 \%$ of DM abundance. 
It is important to note that large LTA can be generated only for small values of $\imw$. This means that the values of the total mixing $|U|^2$ is very close to the see-saw ones.
Interestingly,  Higgs inflation in Einstein-Cartan formulation provides an independent mechanism of DM production~\cite{Shaposhnikov:2020aen}. In this mechanism, the only constraint on the $\nu$MSM parameters comes from BAU, as
no LTA is required. 

We identify further directions.
\begin{itemize}
    \item A regime of the asymmetry generation during freeze-out and decays have been studied in refs.~\cite{Canetti:2012vf,Canetti:2012kh}. There were many significant improvements in understanding of the system since these works have been performed. 
    The recent study~\cite{Ghiglieri:2020ulj} has demonstrated that production of the correct amount of DM in the $\nu$MSM is possible also accounting for all the improvements and effects, such as the entropy dilution. Still, an updated study of the parameter space is missing. 
    \item The possible uncertainties related to the interaction rates and especially the effective Hamiltonian could be large. These points need to be further clarified.
    \item Last but not least, as we have discussed in section~\ref{sec:chiral_asymmetriy_and_magnetic_fields} non-perturbative effects associated with Abelian anomaly may play an important role in the dynamics of the system. Therefore, two important tasks arise: \emph{(i)} derivation of a kinetic equation~\eqref{eff} from the first principles and \emph{(ii)} determination of the parameters $\gamma_C$ and $\kappa$ entering this equations.
\end{itemize}

\acknowledgments
We appreciate fruitful discussions with Jacopo Ghiglieri, Juraj Klaric, and Mikko Laine and thank them for their helpful comments on the draft of this paper. IT thanks Jean-Loup Tastet for help with Julia programming language.
We are grateful to the anonymous referee for many important comments that improved the paper.
This work was supported by the 
ERC-AdG-2015 grant 694896 and by the Swiss National Science Foundation Excellence grant 200020B\underline{ }182864.

\appendix

\section{Late time asymmetries, inverted hierarchy, and CP-asymmetry} 
\label{sec:late_time_asymmetries_and_inverted_hierarchy}
In this section we provide more details on the inverted hierarchy case.
As has been discussed in section~\ref{sec:results}, the asymmetry in the IH case is much smaller than in the NH case. In order to understand this fact, we need to examine the rates entering kinetic equations~\eqref{kin_eq}. Namely, let us consider the HNL washout rate $\Gamma_{N}$. It can be written as
\begin{equation}
    \Gamma_N = \gamma_+ \sum_{\alpha} \left(\begin{array}{cc}
h_{\alpha 3} h_{\alpha 3}^{*} & -h_{\alpha 3} h_{\alpha 2}^{*} \\
-h_{\alpha 2} h_{\alpha 3}^{*} & h_{\alpha 2} h_{\alpha 2}^{*}
\end{array}\right) + \gamma_- \sum_{\alpha} \left(\begin{array}{cc}
h_{\alpha 2} h_{\alpha 2}^{*} & -h_{\alpha 3} h_{\alpha 2}^{*} \\
-h_{\alpha 2} h_{\alpha 3}^{*} & h_{\alpha 3} h_{\alpha 3}^{*}
\end{array}\right),
\label{G_N}
\end{equation}
where  $\gamma_\pm$ are functions of the temperature and momentum, whereas $h_{\alpha I}$ are Yukawas in the flavour basis (see, e.g. ref.~\cite{Eijima:2018qke}). The combinations of Yukawas entering~\eqref{G_N} can be expressed using the parameter $\omega$ as
\begin{equation}
\begin{aligned}
\sum_{\alpha} h_{\alpha 2} h_{\alpha 2}^{*} &=\frac{\left(m_{h}+m_{l}\right) M}{2 v^{2}}  \exp( - 2 \imw) \\
\sum_{\alpha} h_{\alpha 3} h_{\alpha 3}^{*} &=\frac{\left(m_{h}+m_{l}\right) M}{2 v^{2}} \exp(2 \imw)  \\
\sum_{\alpha} \Re\left(h_{\alpha 2} h_{\alpha 3}^{*}\right) &=\frac{\left(m_{h}-m_{l}\right) M}{2 v^{2}} \cos (2 \Re \omega) \\
\sum_{\alpha} \operatorname{Im}\left(h_{\alpha 2} h_{\alpha 3}^{*}\right) &=\frac{\left(m_{h}-m_{l}\right) M}{2 v^{2}} \sin (2 \Re \omega),
\end{aligned}
\label{yukawa_comb}
\end{equation}
where $v = 173.1$~GeV is the Higgs vev, $m_h$ is the mass of the heaviest active neutrino, and $m_l$ is the mass of the lighter one.
\begin{equation}
    \begin{aligned}
    m_{h}+m_{l} & \stackrel{\text{NH}}{=} m_{atm} + m_{sol} 
    \simeq 5.86\times10^{-11}~\text{GeV},\\
    m_{h}+m_{l} & \stackrel{\text{IH}}{=} m_{atm} + \sqrt{m_{atm}^2 - m_{sol}^2 }
    \simeq 9.86\times10^{-11}~\text{GeV},\\
    m_{h}-m_{l} & \stackrel{\text{NH}}{=} m_{atm} - m_{sol}
    \simeq 4.13\times10^{-11}~\text{GeV},\\
    m_{h}-m_{l} & \stackrel{\text{IH}}{=} m_{atm} - \sqrt{m_{atm}^2 - m_{sol}^2 }
    \simeq 7.51\times10^{-13}~\text{GeV}.
\end{aligned}
\label{mass_comb}
\end{equation}
As we have discussed in section~\ref{sub:most_efficient_asymmetry_generation}, the asymmetry peaks at $\imw \simeq 0$. In this case the diagonal entries in~\eqref{G_N} are equal. 
As a consequence, the difference between the eigenvalues of $\Gamma_N$ is proportional to the magnitude of the off-diagonal elements.
We can see from the expressions above that the off-diagonal entries of $\Gamma_N$ are 
$\sim 55$ times smaller in the IH case. 
In physical terms it means that the equilibration rates of two HNLs are practically the same in the IH case (cf figures 2 and 4 in ref.~\cite{Eijima:2017anv}). 
This implies that less asymmetry can build up between the moments when two HNLs enter equilibrium, since it happens almost simultaneously.
In the NH case the equilibration rates are different, and therefore more asymmetry can survive in the HNL sector.

Yet another consequence of eq.~\eqref{yukawa_comb} is related to the $CP$ violation in the oscillations of HNLs. 
The effective Hamiltonian $H_N$ describing these oscillations depends on the same combinations of Yukawas as~\eqref{G_N}. 
The $CP$ violating effects are associated with the imaginary part of $H_N$ which---as one can see from eqs.~\eqref{yukawa_comb} and \eqref{mass_comb}---is also smaller in the IH case.


\bibliographystyle{JHEP}
\bibliography{refs_LTA}

\providecommand{\href}[2]{#2}\begingroup\raggedright\begin{thebibliography}{100}

\bibitem{Asaka:2005an}
T.~Asaka, S.~Blanchet and M.~Shaposhnikov, \emph{{The $\nu$MSM, dark matter and
  neutrino masses}},
  \href{https://doi.org/10.1016/j.physletb.2005.09.070}{\emph{Phys. Lett. B}
  {\bfseries 631} (2005) 151--156},
  [\href{https://arxiv.org/abs/hep-ph/0503065}{{\ttfamily hep-ph/0503065}}].

\bibitem{Asaka:2005pn}
T.~Asaka and M.~Shaposhnikov, \emph{{The $\nu$MSM, dark matter and baryon
  asymmetry of the universe}},
  \href{https://doi.org/10.1016/j.physletb.2005.06.020}{\emph{Phys. Lett. B}
  {\bfseries 620} (2005) 17--26},
  [\href{https://arxiv.org/abs/hep-ph/0505013}{{\ttfamily hep-ph/0505013}}].

\bibitem{Dodelson:1993je}
S.~Dodelson and L.~M. Widrow, \emph{{Sterile-neutrinos as dark matter}},
  \href{https://doi.org/10.1103/PhysRevLett.72.17}{\emph{Phys. Rev. Lett.}
  {\bfseries 72} (1994) 17--20},
  [\href{https://arxiv.org/abs/hep-ph/9303287}{{\ttfamily hep-ph/9303287}}].

\bibitem{Shi:1998km}
X.-D. Shi and G.~M. Fuller, \emph{{A New dark matter candidate: Nonthermal
  sterile neutrinos}},
  \href{https://doi.org/10.1103/PhysRevLett.82.2832}{\emph{Phys. Rev. Lett.}
  {\bfseries 82} (1999) 2832--2835},
  [\href{https://arxiv.org/abs/astro-ph/9810076}{{\ttfamily
  astro-ph/9810076}}].

\bibitem{Abazajian:2001nj}
K.~Abazajian, G.~M. Fuller and M.~Patel, \emph{{Sterile neutrino hot, warm, and
  cold dark matter}},
  \href{https://doi.org/10.1103/PhysRevD.64.023501}{\emph{Phys. Rev. D}
  {\bfseries 64} (2001) 023501},
  [\href{https://arxiv.org/abs/astro-ph/0101524}{{\ttfamily
  astro-ph/0101524}}].

\bibitem{Asaka:2006nq}
T.~Asaka, M.~Laine and M.~Shaposhnikov, \emph{{Lightest sterile neutrino
  abundance within the $\nu$MSM}},
  \href{https://doi.org/10.1088/1126-6708/2007/01/091}{\emph{JHEP} {\bfseries
  01} (2007) 091}, [\href{https://arxiv.org/abs/hep-ph/0612182}{{\ttfamily
  hep-ph/0612182}}].

\bibitem{Laine:2008pg}
M.~Laine and M.~Shaposhnikov, \emph{{Sterile neutrino dark matter as a
  consequence of $\nu$MSM-induced lepton asymmetry}},
  \href{https://doi.org/10.1088/1475-7516/2008/06/031}{\emph{JCAP} {\bfseries
  06} (2008) 031}, [\href{https://arxiv.org/abs/0804.4543}{{\ttfamily
  0804.4543}}].

\bibitem{Shaposhnikov:2020aen}
M.~Shaposhnikov, A.~Shkerin, I.~Timiryasov and S.~Zell, \emph{{Einstein-Cartan
  Portal to Dark Matter}},
  \href{https://doi.org/10.1103/PhysRevLett.127.169901}{\emph{Phys. Rev. Lett.}
  {\bfseries 126} (2021) 161301},
  [\href{https://arxiv.org/abs/2008.11686}{{\ttfamily 2008.11686}}].

\bibitem{Minkowski:1977sc}
P.~Minkowski, \emph{{$\mu \to e\gamma$ at a Rate of One Out of $10^{9}$ Muon
  Decays?}}, \href{https://doi.org/10.1016/0370-2693(77)90435-X}{\emph{Phys.
  Lett. B} {\bfseries 67} (1977) 421--428}.

\bibitem{Yanagida:1979as}
T.~Yanagida, \emph{{Horizontal Gauge Symmetry and Masses of Neutrinos}},
  {\emph{Conf. Proc. C} {\bfseries 7902131} (1979) 95--99}.

\bibitem{GellMann:1980vs}
M.~Gell-Mann, P.~Ramond and R.~Slansky, \emph{{Complex Spinors and Unified
  Theories}}, {\emph{Conf. Proc. C} {\bfseries 790927} (1979) 315--321},
  [\href{https://arxiv.org/abs/1306.4669}{{\ttfamily 1306.4669}}].

\bibitem{Mohapatra:1979ia}
R.~N. Mohapatra and G.~Senjanovic, \emph{{Neutrino Mass and Spontaneous Parity
  Nonconservation}},
  \href{https://doi.org/10.1103/PhysRevLett.44.912}{\emph{Phys. Rev. Lett.}
  {\bfseries 44} (1980) 912}.

\bibitem{Schechter:1980gr}
J.~Schechter and J.~Valle, \emph{{Neutrino Masses in SU(2) x U(1) Theories}},
  \href{https://doi.org/10.1103/PhysRevD.22.2227}{\emph{Phys. Rev. D}
  {\bfseries 22} (1980) 2227}.

\bibitem{Schechter:1981cv}
J.~Schechter and J.~Valle, \emph{{Neutrino Decay and Spontaneous Violation of
  Lepton Number}}, \href{https://doi.org/10.1103/PhysRevD.25.774}{\emph{Phys.
  Rev. D} {\bfseries 25} (1982) 774}.

\bibitem{Akhmedov:1998qx}
E.~K. Akhmedov, V.~Rubakov and A.~Smirnov, \emph{{Baryogenesis via neutrino
  oscillations}},
  \href{https://doi.org/10.1103/PhysRevLett.81.1359}{\emph{Phys. Rev. Lett.}
  {\bfseries 81} (1998) 1359--1362},
  [\href{https://arxiv.org/abs/hep-ph/9803255}{{\ttfamily hep-ph/9803255}}].

\bibitem{Shaposhnikov:2008pf}
M.~Shaposhnikov, \emph{{The $\nu$MSM, leptonic asymmetries, and properties of
  singlet fermions}},
  \href{https://doi.org/10.1088/1126-6708/2008/08/008}{\emph{JHEP} {\bfseries
  08} (2008) 008}, [\href{https://arxiv.org/abs/0804.4542}{{\ttfamily
  0804.4542}}].

\bibitem{Canetti:2012kh}
L.~Canetti, M.~Drewes, T.~Frossard and M.~Shaposhnikov, \emph{{Dark Matter,
  Baryogenesis and Neutrino Oscillations from Right Handed Neutrinos}},
  \href{https://doi.org/10.1103/PhysRevD.87.093006}{\emph{Phys. Rev. D}
  {\bfseries 87} (2013) 093006},
  [\href{https://arxiv.org/abs/1208.4607}{{\ttfamily 1208.4607}}].

\bibitem{Bezrukov:2011sz}
F.~Bezrukov, D.~Gorbunov and M.~Shaposhnikov, \emph{{Late and early time
  phenomenology of Higgs-dependent cutoff}},
  \href{https://doi.org/10.1088/1475-7516/2011/10/001}{\emph{JCAP} {\bfseries
  10} (2011) 001}, [\href{https://arxiv.org/abs/1106.5019}{{\ttfamily
  1106.5019}}].

\bibitem{Boyarsky:2018tvu}
A.~Boyarsky, M.~Drewes, T.~Lasserre, S.~Mertens and O.~Ruchayskiy,
  \emph{{Sterile neutrino Dark Matter}},
  \href{https://doi.org/10.1016/j.ppnp.2018.07.004}{\emph{Prog. Part. Nucl.
  Phys.} {\bfseries 104} (2019) 1--45},
  [\href{https://arxiv.org/abs/1807.07938}{{\ttfamily 1807.07938}}].

\bibitem{Shaposhnikov:2006nn}
M.~Shaposhnikov, \emph{{A Possible symmetry of the $\nu$MSM}},
  \href{https://doi.org/10.1016/j.nuclphysb.2006.11.003}{\emph{Nucl. Phys. B}
  {\bfseries 763} (2007) 49--59},
  [\href{https://arxiv.org/abs/hep-ph/0605047}{{\ttfamily hep-ph/0605047}}].

\bibitem{Canetti:2010aw}
L.~Canetti and M.~Shaposhnikov, \emph{{Baryon Asymmetry of the Universe in the
  $\nu$MSM}}, \href{https://doi.org/10.1088/1475-7516/2010/09/001}{\emph{JCAP}
  {\bfseries 09} (2010) 001},
  [\href{https://arxiv.org/abs/1006.0133}{{\ttfamily 1006.0133}}].

\bibitem{Asaka:2010kk}
T.~Asaka and H.~Ishida, \emph{{Flavour Mixing of Neutrinos and Baryon Asymmetry
  of the Universe}},
  \href{https://doi.org/10.1016/j.physletb.2010.07.016}{\emph{Phys. Lett. B}
  {\bfseries 692} (2010) 105--113},
  [\href{https://arxiv.org/abs/1004.5491}{{\ttfamily 1004.5491}}].

\bibitem{Anisimov:2010gy}
A.~Anisimov, D.~Besak and D.~Bodeker, \emph{{Thermal production of relativistic
  Majorana neutrinos: Strong enhancement by multiple soft scattering}},
  \href{https://doi.org/10.1088/1475-7516/2011/03/042}{\emph{JCAP} {\bfseries
  03} (2011) 042}, [\href{https://arxiv.org/abs/1012.3784}{{\ttfamily
  1012.3784}}].

\bibitem{Asaka:2011wq}
T.~Asaka, S.~Eijima and H.~Ishida, \emph{{Kinetic Equations for Baryogenesis
  via Sterile Neutrino Oscillation}},
  \href{https://doi.org/10.1088/1475-7516/2012/02/021}{\emph{JCAP} {\bfseries
  02} (2012) 021}, [\href{https://arxiv.org/abs/1112.5565}{{\ttfamily
  1112.5565}}].

\bibitem{Besak:2012qm}
D.~Besak and D.~Bodeker, \emph{{Thermal production of ultrarelativistic
  right-handed neutrinos: Complete leading-order results}},
  \href{https://doi.org/10.1088/1475-7516/2012/03/029}{\emph{JCAP} {\bfseries
  03} (2012) 029}, [\href{https://arxiv.org/abs/1202.1288}{{\ttfamily
  1202.1288}}].

\bibitem{Canetti:2012vf}
L.~Canetti, M.~Drewes and M.~Shaposhnikov, \emph{{Sterile Neutrinos as the
  Origin of Dark and Baryonic Matter}},
  \href{https://doi.org/10.1103/PhysRevLett.110.061801}{\emph{Phys. Rev. Lett.}
  {\bfseries 110} (2013) 061801},
  [\href{https://arxiv.org/abs/1204.3902}{{\ttfamily 1204.3902}}].

\bibitem{Drewes:2012ma}
M.~Drewes and B.~Garbrecht, \emph{{Leptogenesis from a GeV Seesaw without Mass
  Degeneracy}}, \href{https://doi.org/10.1007/JHEP03(2013)096}{\emph{JHEP}
  {\bfseries 03} (2013) 096},
  [\href{https://arxiv.org/abs/1206.5537}{{\ttfamily 1206.5537}}].

\bibitem{Shuve:2014zua}
B.~Shuve and I.~Yavin, \emph{{Baryogenesis through Neutrino Oscillations: A
  Unified Perspective}},
  \href{https://doi.org/10.1103/PhysRevD.89.075014}{\emph{Phys. Rev. D}
  {\bfseries 89} (2014) 075014},
  [\href{https://arxiv.org/abs/1401.2459}{{\ttfamily 1401.2459}}].

\bibitem{Bodeker:2014hqa}
D.~Bodeker and M.~Laine, \emph{{Kubo relations and radiative corrections for
  lepton number washout}},
  \href{https://doi.org/10.1088/1475-7516/2014/05/041}{\emph{JCAP} {\bfseries
  05} (2014) 041}, [\href{https://arxiv.org/abs/1403.2755}{{\ttfamily
  1403.2755}}].

\bibitem{Abada:2015rta}
A.~Abada, G.~Arcadi, V.~Domcke and M.~Lucente, \emph{{Lepton number violation
  as a key to low-scale leptogenesis}},
  \href{https://doi.org/10.1088/1475-7516/2015/11/041}{\emph{JCAP} {\bfseries
  11} (2015) 041}, [\href{https://arxiv.org/abs/1507.06215}{{\ttfamily
  1507.06215}}].

\bibitem{Hernandez:2015wna}
P.~Hern\'andez, M.~Kekic, J.~L\'opez-Pav\'on, J.~Racker and N.~Rius,
  \emph{{Leptogenesis in GeV scale seesaw models}},
  \href{https://doi.org/10.1007/JHEP10(2015)067}{\emph{JHEP} {\bfseries 10}
  (2015) 067}, [\href{https://arxiv.org/abs/1508.03676}{{\ttfamily
  1508.03676}}].

\bibitem{Ghiglieri:2016xye}
J.~Ghiglieri and M.~Laine, \emph{{Neutrino dynamics below the electroweak
  crossover}}, \href{https://doi.org/10.1088/1475-7516/2016/07/015}{\emph{JCAP}
  {\bfseries 07} (2016) 015},
  [\href{https://arxiv.org/abs/1605.07720}{{\ttfamily 1605.07720}}].

\bibitem{Hambye:2016sby}
T.~Hambye and D.~Teresi, \emph{{Higgs doublet decay as the origin of the baryon
  asymmetry}},
  \href{https://doi.org/10.1103/PhysRevLett.117.091801}{\emph{Phys. Rev. Lett.}
  {\bfseries 117} (2016) 091801},
  [\href{https://arxiv.org/abs/1606.00017}{{\ttfamily 1606.00017}}].

\bibitem{Hambye:2017elz}
T.~Hambye and D.~Teresi, \emph{{Baryogenesis from L-violating Higgs-doublet
  decay in the density-matrix formalism}},
  \href{https://doi.org/10.1103/PhysRevD.96.015031}{\emph{Phys. Rev. D}
  {\bfseries 96} (2017) 015031},
  [\href{https://arxiv.org/abs/1705.00016}{{\ttfamily 1705.00016}}].

\bibitem{Drewes:2016lqo}
M.~Drewes and S.~Eijima, \emph{{Neutrinoless double $\beta$ decay and low scale
  leptogenesis}},
  \href{https://doi.org/10.1016/j.physletb.2016.09.054}{\emph{Phys. Lett. B}
  {\bfseries 763} (2016) 72--79},
  [\href{https://arxiv.org/abs/1606.06221}{{\ttfamily 1606.06221}}].

\bibitem{Asaka:2016zib}
T.~Asaka, S.~Eijima and H.~Ishida, \emph{{On neutrinoless double beta decay in
  the $\nu$MSM}},
  \href{https://doi.org/10.1016/j.physletb.2016.09.044}{\emph{Phys. Lett. B}
  {\bfseries 762} (2016) 371--375},
  [\href{https://arxiv.org/abs/1606.06686}{{\ttfamily 1606.06686}}].

\bibitem{Drewes:2016gmt}
M.~Drewes, B.~Garbrecht, D.~Gueter and J.~Klaric, \emph{{Leptogenesis from
  Oscillations of Heavy Neutrinos with Large Mixing Angles}},
  \href{https://doi.org/10.1007/JHEP12(2016)150}{\emph{JHEP} {\bfseries 12}
  (2016) 150}, [\href{https://arxiv.org/abs/1606.06690}{{\ttfamily
  1606.06690}}].

\bibitem{Hernandez:2016kel}
P.~Hern\'andez, M.~Kekic, J.~L\'opez-Pav\'on, J.~Racker and J.~Salvado,
  \emph{{Testable Baryogenesis in Seesaw Models}},
  \href{https://doi.org/10.1007/JHEP08(2016)157}{\emph{JHEP} {\bfseries 08}
  (2016) 157}, [\href{https://arxiv.org/abs/1606.06719}{{\ttfamily
  1606.06719}}].

\bibitem{Drewes:2016jae}
M.~Drewes, B.~Garbrecht, D.~Gueter and J.~Klaric, \emph{{Testing the low scale
  seesaw and leptogenesis}},
  \href{https://doi.org/10.1007/JHEP08(2017)018}{\emph{JHEP} {\bfseries 08}
  (2017) 018}, [\href{https://arxiv.org/abs/1609.09069}{{\ttfamily
  1609.09069}}].

\bibitem{Asaka:2017rdj}
T.~Asaka, S.~Eijima, H.~Ishida, K.~Minogawa and T.~Yoshii, \emph{{Initial
  condition for baryogenesis via neutrino oscillation}},
  \href{https://doi.org/10.1103/PhysRevD.96.083010}{\emph{Phys. Rev. D}
  {\bfseries 96} (2017) 083010},
  [\href{https://arxiv.org/abs/1704.02692}{{\ttfamily 1704.02692}}].

\bibitem{Eijima:2017anv}
S.~Eijima and M.~Shaposhnikov, \emph{{Fermion number violating effects in low
  scale leptogenesis}},
  \href{https://doi.org/10.1016/j.physletb.2017.05.068}{\emph{Phys. Lett. B}
  {\bfseries 771} (2017) 288--296},
  [\href{https://arxiv.org/abs/1703.06085}{{\ttfamily 1703.06085}}].

\bibitem{Ghiglieri:2017gjz}
J.~Ghiglieri and M.~Laine, \emph{{GeV-scale hot sterile neutrino oscillations:
  a derivation of evolution equations}},
  \href{https://doi.org/10.1007/JHEP05(2017)132}{\emph{JHEP} {\bfseries 05}
  (2017) 132}, [\href{https://arxiv.org/abs/1703.06087}{{\ttfamily
  1703.06087}}].

\bibitem{Eijima:2017cxr}
S.~Eijima, M.~Shaposhnikov and I.~Timiryasov, \emph{{Freeze-out of baryon
  number in low-scale leptogenesis}},
  \href{https://doi.org/10.1088/1475-7516/2017/11/030}{\emph{JCAP} {\bfseries
  11} (2017) 030}, [\href{https://arxiv.org/abs/1709.07834}{{\ttfamily
  1709.07834}}].

\bibitem{Antusch:2017pkq}
S.~Antusch, E.~Cazzato, M.~Drewes, O.~Fischer, B.~Garbrecht, D.~Gueter et~al.,
  \emph{{Probing Leptogenesis at Future Colliders}},
  \href{https://doi.org/10.1007/JHEP09(2018)124}{\emph{JHEP} {\bfseries 09}
  (2018) 124}, [\href{https://arxiv.org/abs/1710.03744}{{\ttfamily
  1710.03744}}].

\bibitem{Ghiglieri:2017csp}
J.~Ghiglieri and M.~Laine, \emph{{GeV-scale hot sterile neutrino oscillations:
  a numerical solution}},
  \href{https://doi.org/10.1007/JHEP02(2018)078}{\emph{JHEP} {\bfseries 02}
  (2018) 078}, [\href{https://arxiv.org/abs/1711.08469}{{\ttfamily
  1711.08469}}].

\bibitem{Eijima:2018qke}
S.~Eijima, M.~Shaposhnikov and I.~Timiryasov, \emph{{Parameter space of
  baryogenesis in the $\nu$MSM}},
  \href{https://doi.org/10.1007/JHEP07(2019)077}{\emph{JHEP} {\bfseries 07}
  (2019) 077}, [\href{https://arxiv.org/abs/1808.10833}{{\ttfamily
  1808.10833}}].

\bibitem{Ghiglieri:2018wbs}
J.~Ghiglieri and M.~Laine, \emph{{Precision study of GeV-scale resonant
  leptogenesis}}, \href{https://doi.org/10.1007/JHEP02(2019)014}{\emph{JHEP}
  {\bfseries 02} (2019) 014},
  [\href{https://arxiv.org/abs/1811.01971}{{\ttfamily 1811.01971}}].

\bibitem{Ghiglieri:2019kbw}
J.~Ghiglieri and M.~Laine, \emph{{Sterile neutrino dark matter via GeV-scale
  leptogenesis?}}, \href{https://doi.org/10.1007/JHEP07(2019)078}{\emph{JHEP}
  {\bfseries 07} (2019) 078},
  [\href{https://arxiv.org/abs/1905.08814}{{\ttfamily 1905.08814}}].

\bibitem{Bodeker:2019rvr}
D.~B\"odeker and D.~Schr\"oder, \emph{{Kinetic equations for sterile neutrinos
  from thermal fluctuations}},
  \href{https://doi.org/10.1088/1475-7516/2020/02/033}{\emph{JCAP} {\bfseries
  02} (2020) 033}, [\href{https://arxiv.org/abs/1911.05092}{{\ttfamily
  1911.05092}}].

\bibitem{Ghiglieri:2020ulj}
J.~Ghiglieri and M.~Laine, \emph{{Sterile neutrino dark matter via coinciding
  resonances}},
  \href{https://doi.org/10.1088/1475-7516/2020/07/012}{\emph{JCAP} {\bfseries
  07} (2020) 012}, [\href{https://arxiv.org/abs/2004.10766}{{\ttfamily
  2004.10766}}].

\bibitem{Klaric:2020lov}
J.~Klari\'c, M.~Shaposhnikov and I.~Timiryasov, \emph{{Uniting low-scale
  leptogeneses}},  \href{https://arxiv.org/abs/2008.13771}{{\ttfamily
  2008.13771}}.

\bibitem{Domcke:2020ety}
V.~Domcke, M.~Drewes, M.~Hufnagel and M.~Lucente, \emph{{Mev-Scale Seesaw and
  Leptogenesis}}, \href{https://doi.org/10.1007/JHEP01(2021)200}{\emph{JHEP}
  {\bfseries 01} (2021) 200},
  [\href{https://arxiv.org/abs/2009.11678}{{\ttfamily 2009.11678}}].

\bibitem{Liventsev:2013zz}
{\scshape Belle} collaboration, D.~Liventsev et~al., \emph{{Search for heavy
  neutrinos at Belle}},
  \href{https://doi.org/10.1103/PhysRevD.87.071102}{\emph{Phys. Rev. D}
  {\bfseries 87} (2013) 071102},
  [\href{https://arxiv.org/abs/1301.1105}{{\ttfamily 1301.1105}}].

\bibitem{Aaij:2014aba}
{\scshape LHCb} collaboration, R.~Aaij et~al., \emph{{Search for Majorana
  neutrinos in $B^- \to \pi^+\mu^-\mu^-$ decays}},
  \href{https://doi.org/10.1103/PhysRevLett.112.131802}{\emph{Phys. Rev. Lett.}
  {\bfseries 112} (2014) 131802},
  [\href{https://arxiv.org/abs/1401.5361}{{\ttfamily 1401.5361}}].

\bibitem{Artamonov:2014urb}
{\scshape E949} collaboration, A.~Artamonov et~al., \emph{{Search for heavy
  neutrinos in $K^+\to\mu^+\nu_H$ decays}},
  \href{https://doi.org/10.1103/PhysRevD.91.052001}{\emph{Phys. Rev. D}
  {\bfseries 91} (2015) 052001},
  [\href{https://arxiv.org/abs/1411.3963}{{\ttfamily 1411.3963}}].

\bibitem{Aad:2015xaa}
{\scshape ATLAS} collaboration, G.~Aad et~al., \emph{{Search for heavy Majorana
  neutrinos with the ATLAS detector in pp collisions at $ \sqrt{s}=8 $ TeV}},
  \href{https://doi.org/10.1007/JHEP07(2015)162}{\emph{JHEP} {\bfseries 07}
  (2015) 162}, [\href{https://arxiv.org/abs/1506.06020}{{\ttfamily
  1506.06020}}].

\bibitem{Khachatryan:2015gha}
{\scshape CMS} collaboration, V.~Khachatryan et~al., \emph{{Search for heavy
  Majorana neutrinos in $\mu^\pm \mu^\pm+$ jets events in proton-proton
  collisions at $\sqrt{s}$ = 8 TeV}},
  \href{https://doi.org/10.1016/j.physletb.2015.06.070}{\emph{Phys. Lett. B}
  {\bfseries 748} (2015) 144--166},
  [\href{https://arxiv.org/abs/1501.05566}{{\ttfamily 1501.05566}}].

\bibitem{Antusch:2017hhu}
S.~Antusch, E.~Cazzato and O.~Fischer, \emph{{Sterile neutrino searches via
  displaced vertices at LHCb}},
  \href{https://doi.org/10.1016/j.physletb.2017.09.057}{\emph{Phys. Lett. B}
  {\bfseries 774} (2017) 114--118},
  [\href{https://arxiv.org/abs/1706.05990}{{\ttfamily 1706.05990}}].

\bibitem{CortinaGil:2017mqf}
{\scshape NA62} collaboration, E.~Cortina~Gil et~al., \emph{{Search for heavy
  neutral lepton production in $K^+$ decays}},
  \href{https://doi.org/10.1016/j.physletb.2018.01.031}{\emph{Phys. Lett. B}
  {\bfseries 778} (2018) 137--145},
  [\href{https://arxiv.org/abs/1712.00297}{{\ttfamily 1712.00297}}].

\bibitem{Izmaylov:2017lkv}
A.~Izmaylov and S.~Suvorov, \emph{{Search for heavy neutrinos in the ND280 near
  detector of the T2K experiment}},
  \href{https://doi.org/10.1134/S1063779617060223}{\emph{Phys. Part. Nucl.}
  {\bfseries 48} (2017) 984--986}.

\bibitem{Mermod:2017ceo}
{\scshape SHiP} collaboration, P.~Mermod, \emph{{Prospects of the SHiP and NA62
  experiments at CERN for hidden sector searches}},
  \href{https://doi.org/10.22323/1.295.0139}{\emph{PoS} {\bfseries NuFact2017}
  (2017) 139}, [\href{https://arxiv.org/abs/1712.01768}{{\ttfamily
  1712.01768}}].

\bibitem{Drewes:2018gkc}
M.~Drewes, J.~Hajer, J.~Klaric and G.~Lanfranchi, \emph{{NA62 sensitivity to
  heavy neutral leptons in the low scale seesaw model}},
  \href{https://doi.org/10.1007/JHEP07(2018)105}{\emph{JHEP} {\bfseries 07}
  (2018) 105}, [\href{https://arxiv.org/abs/1801.04207}{{\ttfamily
  1801.04207}}].

\bibitem{Ballett:2019bgd}
P.~Ballett, T.~Boschi and S.~Pascoli, \emph{{Heavy Neutral Leptons from
  low-scale seesaws at the DUNE Near Detector}},
  \href{https://doi.org/10.1007/JHEP03(2020)111}{\emph{JHEP} {\bfseries 03}
  (2020) 111}, [\href{https://arxiv.org/abs/1905.00284}{{\ttfamily
  1905.00284}}].

\bibitem{Sirunyan:2018mtv}
{\scshape CMS} collaboration, A.~M. Sirunyan et~al., \emph{{Search for heavy
  neutral leptons in events with three charged leptons in proton-proton
  collisions at $\sqrt{s} =$ 13 TeV}},
  \href{https://doi.org/10.1103/PhysRevLett.120.221801}{\emph{Phys. Rev. Lett.}
  {\bfseries 120} (2018) 221801},
  [\href{https://arxiv.org/abs/1802.02965}{{\ttfamily 1802.02965}}].

\bibitem{SHiP:2018xqw}
{\scshape SHiP} collaboration, C.~Ahdida et~al., \emph{{Sensitivity of the SHiP
  experiment to Heavy Neutral Leptons}},
  \href{https://doi.org/10.1007/JHEP04(2019)077}{\emph{JHEP} {\bfseries 04}
  (2019) 077}, [\href{https://arxiv.org/abs/1811.00930}{{\ttfamily
  1811.00930}}].

\bibitem{Boiarska:2019jcw}
I.~Boiarska, K.~Bondarenko, A.~Boyarsky, S.~Eijima, M.~Ovchynnikov,
  O.~Ruchayskiy et~al., \emph{{Probing Baryon Asymmetry of the Universe at Lhc
  and Ship}},  \href{https://arxiv.org/abs/1902.04535}{{\ttfamily 1902.04535}}.

\bibitem{Bolton:2019pcu}
P.~D. Bolton, F.~F. Deppisch and P.~Bhupal~Dev, \emph{{Neutrinoless double beta
  decay versus other probes of heavy sterile neutrinos}},
  \href{https://doi.org/10.1007/JHEP03(2020)170}{\emph{JHEP} {\bfseries 03}
  (2020) 170}, [\href{https://arxiv.org/abs/1912.03058}{{\ttfamily
  1912.03058}}].

\bibitem{NA62:2020mcv}
{\scshape NA62} collaboration, E.~Cortina~Gil et~al., \emph{{Search for heavy
  neutral lepton production in $K^+$ decays to positrons}},
  \href{https://doi.org/10.1016/j.physletb.2020.135599}{\emph{Phys. Lett. B}
  {\bfseries 807} (2020) 135599},
  [\href{https://arxiv.org/abs/2005.09575}{{\ttfamily 2005.09575}}].

\bibitem{Tastet:2020tzh}
J.-L. Tastet, E.~Goudzovski, I.~Timiryasov and O.~Ruchayskiy, \emph{{Projected
  NA62 sensitivity to heavy neutral lepton production in
  K+\textrightarrow{}\ensuremath{\pi}0e+N decays}},
  \href{https://doi.org/10.1103/PhysRevD.104.055005}{\emph{Phys. Rev. D}
  {\bfseries 104} (2021) 055005},
  [\href{https://arxiv.org/abs/2008.11654}{{\ttfamily 2008.11654}}].

\bibitem{Aad:2019kiz}
{\scshape ATLAS} collaboration, G.~Aad et~al., \emph{{Search for heavy neutral
  leptons in decays of $W$ bosons produced in 13 TeV $pp$ collisions using
  prompt and displaced signatures with the ATLAS detector}},
  \href{https://doi.org/10.1007/JHEP10(2019)265}{\emph{JHEP} {\bfseries 10}
  (2019) 265}, [\href{https://arxiv.org/abs/1905.09787}{{\ttfamily
  1905.09787}}].

\bibitem{Wulz:2019lsz}
{\scshape ATLAS, CMS} collaboration, C.-E. Wulz, \emph{{Techniques and results
  of neutral long-lived particle searches in ATLAS and CMS in LHC Run 2}},  in
  \emph{{54th Rencontres de Moriond on Electroweak Interactions and Unified
  Theories}}, pp.~77--84, 2019,
  \href{https://arxiv.org/abs/1907.13588}{{\ttfamily 1907.13588}}.

\bibitem{Alekhin:2015byh}
S.~Alekhin et~al., \emph{{A facility to Search for Hidden Particles at the CERN
  SPS: the SHiP physics case}},
  \href{https://doi.org/10.1088/0034-4885/79/12/124201}{\emph{Rept. Prog.
  Phys.} {\bfseries 79} (2016) 124201},
  [\href{https://arxiv.org/abs/1504.04855}{{\ttfamily 1504.04855}}].

\bibitem{Curtin:2018mvb}
D.~Curtin et~al., \emph{{Long-Lived Particles at the Energy Frontier: The
  MATHUSLA Physics Case}},
  \href{https://doi.org/10.1088/1361-6633/ab28d6}{\emph{Rept. Prog. Phys.}
  {\bfseries 82} (2019) 116201},
  [\href{https://arxiv.org/abs/1806.07396}{{\ttfamily 1806.07396}}].

\bibitem{Gligorov:2017nwh}
V.~V. Gligorov, S.~Knapen, M.~Papucci and D.~J. Robinson, \emph{{Searching for
  Long-lived Particles: A Compact Detector for Exotics at LHCb}},
  \href{https://doi.org/10.1103/PhysRevD.97.015023}{\emph{Phys. Rev. D}
  {\bfseries 97} (2018) 015023},
  [\href{https://arxiv.org/abs/1708.09395}{{\ttfamily 1708.09395}}].

\bibitem{Feng:2017uoz}
J.~L. Feng, I.~Galon, F.~Kling and S.~Trojanowski, \emph{{ForwArd Search
  ExpeRiment at the LHC}},
  \href{https://doi.org/10.1103/PhysRevD.97.035001}{\emph{Phys. Rev. D}
  {\bfseries 97} (2018) 035001},
  [\href{https://arxiv.org/abs/1708.09389}{{\ttfamily 1708.09389}}].

\bibitem{Kling:2018wct}
F.~Kling and S.~Trojanowski, \emph{{Heavy Neutral Leptons at FASER}},
  \href{https://doi.org/10.1103/PhysRevD.97.095016}{\emph{Phys. Rev. D}
  {\bfseries 97} (2018) 095016},
  [\href{https://arxiv.org/abs/1801.08947}{{\ttfamily 1801.08947}}].

\bibitem{Hirsch:2020klk}
M.~Hirsch and Z.~S. Wang, \emph{{Heavy neutral leptons at ANUBIS}},
  \href{https://doi.org/10.1103/PhysRevD.101.055034}{\emph{Phys. Rev. D}
  {\bfseries 101} (2020) 055034},
  [\href{https://arxiv.org/abs/2001.04750}{{\ttfamily 2001.04750}}].

\bibitem{Bezrukov:2008ut}
F.~Bezrukov, D.~Gorbunov and M.~Shaposhnikov, \emph{{On initial conditions for
  the Hot Big Bang}},
  \href{https://doi.org/10.1088/1475-7516/2009/06/029}{\emph{JCAP} {\bfseries
  06} (2009) 029}, [\href{https://arxiv.org/abs/0812.3622}{{\ttfamily
  0812.3622}}].

\bibitem{planned_paper}
M.~Drewes, J.~Klari\'c, S.~Eijima, M.~Shaposhnikov, I.~Timiryasov and
  Y.~Georis.

\bibitem{Asaka:2011pb}
T.~Asaka, S.~Eijima and H.~Ishida, \emph{{Mixing of Active and Sterile
  Neutrinos}}, \href{https://doi.org/10.1007/JHEP04(2011)011}{\emph{JHEP}
  {\bfseries 04} (2011) 011},
  [\href{https://arxiv.org/abs/1101.1382}{{\ttfamily 1101.1382}}].

\bibitem{Ghiglieri:2015jua}
J.~Ghiglieri and M.~Laine, \emph{{Improved determination of sterile neutrino
  dark matter spectrum}},
  \href{https://doi.org/10.1007/JHEP11(2015)171}{\emph{JHEP} {\bfseries 11}
  (2015) 171}, [\href{https://arxiv.org/abs/1506.06752}{{\ttfamily
  1506.06752}}].

\bibitem{Casas:2001sr}
J.~Casas and A.~Ibarra, \emph{{Oscillating neutrinos and $\mu \to e, \gamma$}},
  \href{https://doi.org/10.1016/S0550-3213(01)00475-8}{\emph{Nucl. Phys. B}
  {\bfseries 618} (2001) 171--204},
  [\href{https://arxiv.org/abs/hep-ph/0103065}{{\ttfamily hep-ph/0103065}}].

\bibitem{Laine:2006cp}
M.~Laine and Y.~Schroder, \emph{{Quark mass thresholds in QCD thermodynamics}},
  \href{https://doi.org/10.1103/PhysRevD.73.085009}{\emph{Phys. Rev. D}
  {\bfseries 73} (2006) 085009},
  [\href{https://arxiv.org/abs/hep-ph/0603048}{{\ttfamily hep-ph/0603048}}].

\bibitem{Laine:2015kra}
M.~Laine and M.~Meyer, \emph{{Standard Model thermodynamics across the
  electroweak crossover}},
  \href{https://doi.org/10.1088/1475-7516/2015/07/035}{\emph{JCAP} {\bfseries
  07} (2015) 035}, [\href{https://arxiv.org/abs/1503.04935}{{\ttfamily
  1503.04935}}].

\bibitem{Scherrer:1984fd}
R.~J. Scherrer and M.~S. Turner, \emph{{Decaying Particles Do Not Heat Up the
  Universe}}, \href{https://doi.org/10.1103/PhysRevD.31.681}{\emph{Phys. Rev.
  D} {\bfseries 31} (1985) 681}.

\bibitem{Asaka:2006ek}
T.~Asaka, M.~Shaposhnikov and A.~Kusenko, \emph{{Opening a new window for warm
  dark matter}},
  \href{https://doi.org/10.1016/j.physletb.2006.05.067}{\emph{Phys. Lett. B}
  {\bfseries 638} (2006) 401--406},
  [\href{https://arxiv.org/abs/hep-ph/0602150}{{\ttfamily hep-ph/0602150}}].

\bibitem{Gorbunov:2007ak}
D.~Gorbunov and M.~Shaposhnikov, \emph{{How to find neutral leptons of the
  $\nu$MSM?}}, \href{https://doi.org/10.1088/1126-6708/2007/10/015}{\emph{JHEP}
  {\bfseries 10} (2007) 015},
  [\href{https://arxiv.org/abs/0705.1729}{{\ttfamily 0705.1729}}].

\bibitem{Bondarenko:2018ptm}
K.~Bondarenko, A.~Boyarsky, D.~Gorbunov and O.~Ruchayskiy, \emph{{Phenomenology
  of GeV-scale Heavy Neutral Leptons}},
  \href{https://doi.org/10.1007/JHEP11(2018)032}{\emph{JHEP} {\bfseries 11}
  (2018) 032}, [\href{https://arxiv.org/abs/1805.08567}{{\ttfamily
  1805.08567}}].

\bibitem{Dolgov:2000jw}
A.~Dolgov, S.~Hansen, G.~Raffelt and D.~Semikoz, \emph{{Heavy sterile
  neutrinos: Bounds from big bang nucleosynthesis and SN1987A}},
  \href{https://doi.org/10.1016/S0550-3213(00)00566-6}{\emph{Nucl. Phys. B}
  {\bfseries 590} (2000) 562--574},
  [\href{https://arxiv.org/abs/hep-ph/0008138}{{\ttfamily hep-ph/0008138}}].

\bibitem{Ruchayskiy:2012si}
O.~Ruchayskiy and A.~Ivashko, \emph{{Restrictions on the lifetime of sterile
  neutrinos from primordial nucleosynthesis}},
  \href{https://doi.org/10.1088/1475-7516/2012/10/014}{\emph{JCAP} {\bfseries
  10} (2012) 014}, [\href{https://arxiv.org/abs/1202.2841}{{\ttfamily
  1202.2841}}].

\bibitem{Gelmini:2020ekg}
G.~B. Gelmini, M.~Kawasaki, A.~Kusenko, K.~Murai and V.~Takhistov, \emph{{Big
  Bang Nucleosynthesis constraints on sterile neutrino and lepton asymmetry of
  the Universe}},
  \href{https://doi.org/10.1088/1475-7516/2020/09/051}{\emph{JCAP} {\bfseries
  09} (2020) 051}, [\href{https://arxiv.org/abs/2005.06721}{{\ttfamily
  2005.06721}}].

\bibitem{Sabti:2020yrt}
N.~Sabti, A.~Magalich and A.~Filimonova, \emph{{An Extended Analysis of Heavy
  Neutral Leptons during Big Bang Nucleosynthesis}},
  \href{https://doi.org/10.1088/1475-7516/2020/11/056}{\emph{JCAP} {\bfseries
  11} (2020) 056}, [\href{https://arxiv.org/abs/2006.07387}{{\ttfamily
  2006.07387}}].

\bibitem{Boyarsky:2020dzc}
A.~Boyarsky, M.~Ovchynnikov, O.~Ruchayskiy and V.~Syvolap, \emph{{Improved Big
  Bang Nucleosynthesis Constraints on Heavy Neutral Leptons}},
  \href{https://doi.org/10.1103/PhysRevD.104.023517}{\emph{Phys. Rev. D}
  {\bfseries 104} (2021) 023517},
  [\href{https://arxiv.org/abs/2008.00749}{{\ttfamily 2008.00749}}].

\bibitem{Roy:2010xq}
A.~Roy and M.~Shaposhnikov, \emph{{Resonant Production of the Sterile Neutrino
  Dark Matter and Fine-Tunings in the [Nu]Msm}},
  \href{https://doi.org/10.1103/PhysRevD.82.056014}{\emph{Phys. Rev. D}
  {\bfseries 82} (2010) 056014},
  [\href{https://arxiv.org/abs/1006.4008}{{\ttfamily 1006.4008}}].

\bibitem{Asaka:2006rw}
T.~Asaka, M.~Laine and M.~Shaposhnikov, \emph{{On the hadronic contribution to
  sterile neutrino production}},
  \href{https://doi.org/10.1088/1126-6708/2006/06/053}{\emph{JHEP} {\bfseries
  06} (2006) 053}, [\href{https://arxiv.org/abs/hep-ph/0605209}{{\ttfamily
  hep-ph/0605209}}].

\bibitem{Bodeker:2020hbo}
D.~Bodeker and A.~Klaus, \emph{{Sterile neutrino dark matter: Impact of
  active-neutrino opacities}},
  \href{https://doi.org/10.1007/JHEP07(2020)218}{\emph{JHEP} {\bfseries 07}
  (2020) 218}, [\href{https://arxiv.org/abs/2005.03039}{{\ttfamily
  2005.03039}}].

\bibitem{LSODE}
K.~Radhakrishnan and A.~C. Hindmarsh, \emph{Description and use of lsode, the
  livermore solver for ordinary differential equations}, {\emph{Lawrence
  Livermore National Laboratory Report} (1993) }.

\bibitem{Klaric:2021cpi}
J.~Klari\'c, M.~Shaposhnikov and I.~Timiryasov, \emph{{Reconciling resonant
  leptogenesis and baryogenesis via neutrino oscillations}},
  \href{https://doi.org/10.1103/PhysRevD.104.055010}{\emph{Phys. Rev. D}
  {\bfseries 104} (2021) 055010},
  [\href{https://arxiv.org/abs/2103.16545}{{\ttfamily 2103.16545}}].

\bibitem{Blanchet:2009kk}
S.~Blanchet, T.~Hambye and F.-X. Josse-Michaux, \emph{{Reconciling leptogenesis
  with observable mu ---\ensuremath{>} e gamma rates}},
  \href{https://doi.org/10.1007/JHEP04(2010)023}{\emph{JHEP} {\bfseries 04}
  (2010) 023}, [\href{https://arxiv.org/abs/0912.3153}{{\ttfamily 0912.3153}}].

\bibitem{Julia-2017}
J.~Bezanson, A.~Edelman, S.~Karpinski and V.~B. Shah, \emph{Julia: A fresh
  approach to numerical computing},
  \href{https://doi.org/10.1137/141000671}{\emph{SIAM {R}eview} {\bfseries 59}
  (2017) 65--98}.

\bibitem{rackauckas2017differentialequations}
C.~Rackauckas and Q.~Nie, \emph{Differentialequations.jl--a performant and
  feature-rich ecosystem for solving differential equations in julia},
  {\emph{Journal of Open Research Software} {\bfseries 5} (2017) }.

\bibitem{Giovannini:1997eg}
M.~Giovannini and M.~Shaposhnikov, \emph{{Primordial hypermagnetic fields and
  triangle anomaly}},
  \href{https://doi.org/10.1103/PhysRevD.57.2186}{\emph{Phys. Rev. D}
  {\bfseries 57} (1998) 2186--2206},
  [\href{https://arxiv.org/abs/hep-ph/9710234}{{\ttfamily hep-ph/9710234}}].

\bibitem{Kamada:2018tcs}
K.~Kamada, \emph{{Return of grand unified theory baryogenesis: Source of
  helical hypermagnetic fields for the baryon asymmetry of the universe}},
  \href{https://doi.org/10.1103/PhysRevD.97.103506}{\emph{Phys. Rev. D}
  {\bfseries 97} (2018) 103506},
  [\href{https://arxiv.org/abs/1802.03055}{{\ttfamily 1802.03055}}].

\bibitem{Joyce:1997uy}
M.~Joyce and M.~E. Shaposhnikov, \emph{{Primordial magnetic fields,
  right-handed electrons, and the Abelian anomaly}},
  \href{https://doi.org/10.1103/PhysRevLett.79.1193}{\emph{Phys. Rev. Lett.}
  {\bfseries 79} (1997) 1193--1196},
  [\href{https://arxiv.org/abs/astro-ph/9703005}{{\ttfamily
  astro-ph/9703005}}].

\bibitem{Boyarsky:2011uy}
A.~Boyarsky, J.~Frohlich and O.~Ruchayskiy, \emph{{Self-consistent evolution of
  magnetic fields and chiral asymmetry in the early Universe}},
  \href{https://doi.org/10.1103/PhysRevLett.108.031301}{\emph{Phys. Rev. Lett.}
  {\bfseries 108} (2012) 031301},
  [\href{https://arxiv.org/abs/1109.3350}{{\ttfamily 1109.3350}}].

\bibitem{Boyarsky:2020cyk}
A.~Boyarsky, V.~Cheianov, O.~Ruchayskiy and O.~Sobol, \emph{{Evolution of the
  Primordial Axial Charge Across Cosmic Times}},
  \href{https://doi.org/10.1103/PhysRevLett.126.021801}{\emph{Phys. Rev. Lett.}
  {\bfseries 126} (2021) 021801},
  [\href{https://arxiv.org/abs/2007.13691}{{\ttfamily 2007.13691}}].

\bibitem{Boyarsky:2020ani}
A.~Boyarsky, V.~Cheianov, O.~Ruchayskiy and O.~Sobol, \emph{{Equilibration of
  the Chiral Asymmetry Due to Finite Electron Mass in Electron-Positron
  Plasma}}, \href{https://doi.org/10.1103/PhysRevD.103.013003}{\emph{Phys. Rev.
  D} {\bfseries 103} (2021) 013003},
  [\href{https://arxiv.org/abs/2008.00360}{{\ttfamily 2008.00360}}].

\bibitem{Figueroa:2017hun}
D.~G. Figueroa and M.~Shaposhnikov, \emph{{Anomalous non-conservation of
  fermion/chiral number in Abelian gauge theories at finite temperature}},
  \href{https://doi.org/10.1007/JHEP04(2018)026}{\emph{JHEP} {\bfseries 04}
  (2018) 026}, [\href{https://arxiv.org/abs/1707.09967}{{\ttfamily
  1707.09967}}].

\bibitem{Figueroa:2019jsi}
D.~G. Figueroa, A.~Florio and M.~Shaposhnikov, \emph{{Chiral charge dynamics in
  Abelian gauge theories at finite temperature}},
  \href{https://doi.org/10.1007/JHEP10(2019)142}{\emph{JHEP} {\bfseries 10}
  (2019) 142}, [\href{https://arxiv.org/abs/1904.11892}{{\ttfamily
  1904.11892}}].

\bibitem{Long:2013tha}
A.~J. Long, E.~Sabancilar and T.~Vachaspati, \emph{{Leptogenesis and Primordial
  Magnetic Fields}},
  \href{https://doi.org/10.1088/1475-7516/2014/02/036}{\emph{JCAP} {\bfseries
  02} (2014) 036}, [\href{https://arxiv.org/abs/1309.2315}{{\ttfamily
  1309.2315}}].

\bibitem{Bodeker:2019ajh}
D.~B\"odeker and D.~Schr\"oder, \emph{{Equilibration of right-handed
  electrons}}, \href{https://doi.org/10.1088/1475-7516/2019/05/010}{\emph{JCAP}
  {\bfseries 05} (2019) 010},
  [\href{https://arxiv.org/abs/1902.07220}{{\ttfamily 1902.07220}}].

\bibitem{Jackson:2019tnr}
G.~Jackson and M.~Laine, \emph{{A thermal neutrino interaction rate at NLO}},
  \href{https://doi.org/10.1016/j.nuclphysb.2019.114870}{\emph{Nucl. Phys. B}
  {\bfseries 950} (2020) 114870},
  [\href{https://arxiv.org/abs/1910.12880}{{\ttfamily 1910.12880}}].

\bibitem{Weldon:1982bn}
H.~Weldon, \emph{{Effective Fermion Masses of Order gT in High Temperature
  Gauge Theories with Exact Chiral Invariance}},
  \href{https://doi.org/10.1103/PhysRevD.26.2789}{\emph{Phys. Rev. D}
  {\bfseries 26} (1982) 2789}.

\end{thebibliography}\endgroup

\end{document}